\documentclass{aastex631}
\usepackage{newtxtext,newtxmath}
\shorttitle{Less effective hydrodynamic escape of H$_{2}$-H$_{2}$O atmospheres}
\shortauthors{Yoshida et al.}
\graphicspath{{./}{figures/}}

\begin{document}

\title{Less effective hydrodynamic escape of H$_{2}$-H$_{2}$O atmospheres on terrestrial planets orbiting pre-main sequence M dwarfs}

\author{Tatsuya Yoshida}
\affiliation{Faculty of Science, Tohoku University, Sendai, Miyagi 980-8578, Japan}

\author{Naoki Terada}
\affiliation{Faculty of Science, Tohoku University, Sendai, Miyagi 980-8578, Japan}

\author{Masahiro Ikoma}
\affiliation{Division of Science, National Astronomical Observatory of Japan, Mitaka, Tokyo 181-8588, Japan}

\author{Kiyoshi Kuramoto}
\affiliation{Faculty of Science, Hokkaido University, Sapporo, Hokkaido 060-0810, Japan}



\begin{abstract}
	Terrestrial planets currently in the habitable zones around M dwarfs likely experienced a long-term runaway greenhouse condition because of a slow decline in host-stellar luminosity in its pre-main sequence phase. Accordingly, they might have lost significant portions of their atmospheres including water vapor at high concentration by hydrodynamic escape induced by the strong stellar XUV irradiation. However, the atmospheric escape rates remain highly uncertain due partly to a lack of understanding of the effect of radiative cooling in the escape outflows. Here we carry out 1-D hydrodynamic escape simulations for an H$_{2}$-H$_{2}$O atmosphere on a planet with mass of $1M_{\oplus}$ considering radiative and chemical processes to estimate the atmospheric escape rate and follow the atmospheric evolution during the early runaway greenhouse phase. We find that the atmospheric escape rate decreases with the basal H$_{2}$O/H$_{2}$ ratio due to the energy loss by the radiative cooling of H$_{2}$O and chemical products such as OH and H$_{3}^{+}$: the escape rate of H$_{2}$ becomes one order of magnitude smaller when the basal H$_{2}$O/H$_{2}=0.1$ than that of the pure hydrogen atmosphere. The timescale for H$_{2}$ escape exceeds the duration of the early runaway greenhouse phase, depending on the initial atmospheric amount and composition, indicating that H$_{2}$ and H$_{2}$O could be left behind after the end of the runaway greenhouse phase. Our results suggest that temperate and reducing environments with oceans could be formed on some terrestrial planets around M dwarfs.
\end{abstract}



\section{Introduction} \label{sec:intro}
	Planets around M dwarfs are thought to have difficulty in becoming habitable. During the pre-main sequence phase, M dwarfs are theoretically predicted to be 1 or even 2 orders of magnitude more luminous than when they reach the main sequence, and take up to 1 Gyr to settle onto the main sequence (Baraffe et al., 1998). Thus, planets currently in the habitable zones in the main sequence phase are likely to have been in runaway greenhouse conditions, provided they had sufficient surface water (e.g., Luger and Barnes, 2015). Moreover, XUV irradiation in the habitable zones around M dwarfs is estimated to be much stronger than that around Sun-like stars (e.g., Wheatley et al. 2017). Accordingly, terrestrial planets around M dwarfs likely lose significant portions of their atmospheres including water vapor at high concentration by hydrodynamic escape during the early runaway greenhouse phase. Previous studies demonstrated that such planets lose water with the amount equivalent to that in several Earth's oceans by hydrodynamic escape and some O$_{2}$ accumulates as residue depending on the XUV flux and planetary mass (e.g., Luger and Barnes, 2015). Such massive water loss and accumulation of oxygen have negative effects on the origin of life in that liquid water is likely essential for life and highly oxidized environments are unsuitable for the formation of important prebiotic organic molecules (e.g., Schlesinger and Miller, 1983).
	
	Previous studies on atmospheric hydrodynamic escape for terrestrial planets orbiting M dwarfs considered mainly pure H$_{2}$O atmospheres (Luger and Barnes, 2015; Tian, 2015; Tian and Ida, 2015; Bolmont et al. 2017; Bourrier et al. 2017; Johnstone, 2020) or nebula-captured H$_{2}$-dominated atmospheres (Luger et al., 2015; Owen and Mohanty, 2016; Hori and Ogihara, 2020). On the other hand, planet formation theories suggest that accreting terrestrial planets have mixed atmospheres of H$_{2}$ and H$_{2}$O through impact degassing from planetary building blocks and gravitational capture of the surrounding nebular gas. If the conditions are right, water vapor is produced from nebula-captured atmospheres through oxidation of hydrogen by oxides at the surface (Ikoma and Genda, 2006; Kimura and Ikoma, 2020), and on the contrary, hydrogen is produced from impact-generated water vapor through chemical reduction by metallic iron in building blocks (Kuramoto and Matsui, 1996).
	
	The effect of the radiative cooling by H$_{2}$O and chemical products from H$_{2}$ and H$_{2}$O on the hydrodynamic escape has not been fully investigated. Johnstone (2020) considered the radiative emission of H$_{2}$O in rotational bands in the escape outflows of H$_{2}$O atmospheres and indicated that the radiative cooling may have little effect on the hydrodynamic escape. However, Johnstone (2020) neglected the emission in vibrational bands: the effect of the radiative cooling by H$_{2}$O is expected to be larger when the emission in vibrational bands is taken into account because its intensity is higher than that in rotational bands in part of the temperature range of the outflows (e.g., Rothman et al., 2013). Moreover, radiatively active chemical products such as OH and H$_{3}^{+}$ are known to enhance the effect of the radiative cooling (Yoshida and Kuramoto, 2021). If their radiative cooling suppresses the hydrodynamic escape sufficiently, the timescale for H$_{2}$ escape is prolonged compared to that for pure hydrogen atmospheres, and the amounts of water loss and accumulated oxygen are suppressed compared to the previous estimations. 
	
	In this study, we apply our 1-D hydrodynamic escape model of multi-component atmospheres considering radiative and chemical processes developed by Yoshida and Kuramoto (2020) to H$_{2}$-H$_{2}$O atmospheres on terrestrial planets with mass of $1M_{\oplus}$ around pre-main sequence M dwarfs and estimate their atmospheric escape rates. Here we consider line emission in a wide range of wavelength by radiatively active species including chemical products in the escape outflows. Then, we propose possible evolutionary tracks of H$_{2}$-H$_{2}$O atmospheres in the runaway greenhouse state during the host-stellar pre-main sequence phase. This paper is organized as follows. In Section 2, we describe the outline of our hydrodynamic escape model. In Section 3, we show the numerical results of the atmospheric profiles, energy balance, and atmospheric escape rate. In Section 4, we discuss the difference of our results from those of Johnstone (2020) and possible atmospheric evolutionary tracks estimated from the calculated escape rate.

\section{Model description} \label{sec:model}
	A radially one-dimensional hydrodynamic escape model developed by Yoshida \& Kuramoto (2020) is applied with some modifications to the chemical  and radiative processes. The detail of the model is described in Appendix A and Yoshida \& Kuramoto (2020). Below, we present the outline of the model.
	
	We suppose a rocky planet with mass of $1M_{\oplus}$ orbiting at 0.02 au around an M8 dwarf with mass of 0.09 $M_{\odot}$, the stellar properties of which are similar to those of TRAPPIST-1 (Grootel et al., 2018). The orbit of 0.02 au corresponds to the inner edge of the habitable zone during the main sequence phase of such an M8 dwarf (Kopparapu et al., 2013). The dependence of the atmospheric escape on the orbital radius is discussed later.
	
	The gas at the bottom of the atmosphere is assumed to be composed of H$_{2}$ and H$_{2}$O. For simplicity, other components such as carbon species and nitrogen species are neglected because H$_{2}$O would dominate the atmospheric molecular species bearing heavy elements under the runaway greenhouse state referring to the volatile compositions of Earth and volatile-rich primitive meteorites, in which the amount of H$_{2}$O is larger than those of carbon and nitrogen (e.g., Marty, 2012). The lower boundary is set at $r=R_{p}+1000\,\mathrm{km}$ ($\equiv r_{0}$), where $R_{p}$ is the radius of the planet. The number density of H$_{2}$ at the lower boundary is set at $1\times 10^{19}\,\mathrm{m^{-3}}$ assuming the typical gas density at the altitude above which the stellar XUV irradiation is absorbed completely (e.g., Kasting et al., 1983). The number density of H$_{2}$O is given as a parameter in each simulation run. In each simulation run, the atmospheric density and temperature at the lower boundary are fixed. The temperature at the lower boundary is set at 400 K, corresponding to the skin temperature, which is the asymptotic temperature at high altitudes of the upper atmosphere. The upper boundary is set at $r=50\,R_{p}$. The other physical quantities at the lower and upper boundaries are estimated by linear extrapolations from the calculated domain.

	The basic equations in this model are the fluid equations of continuity, momentum, and energy for a multi-component gas considering chemical and radiative processes (Appendix A1). They are solved by numerical integration about time until the physical quantities settle into their steady profiles by the same method that Yoshida \& Kuramoto (2020) used (see also Appendix A4).	

	As for the chemical processes, 93 chemical reactions are considered for 15 atmospheric components: H$_{2}$, H$_{2}$O, H, O, O($^{1}$D), OH, O$_{2}$, H$^{+}$, H$_{2}^{+}$, H$_{3}^{+}$, O$^{+}$, O$_{2}^{+}$, OH$^{+}$, H$_{2}$O$^{+}$, and H$_{3}$O$^{+}$ (Table A1, A2). See Appendix A2 for the model formulation of chemical processes.

	This study adopts the X-ray and UV spectrum profile from 0.1 to 280 nm estimated for the M8 dwarf TRAPPIST-1 (Peacock et al., 2019). The present total XUV flux is $1.78\times 10^{-1}\,\mathrm{J\,m^{-2}\,s^{-1}}$ at 0.02 au. The XUV luminosity during the pre-main sequence phase is assumed to be 0.001 times the total stellar luminosity, which is about 10-230 times the present when the stellar age is from 10 Myr to 1 Gyr (Luger and Barnes, 2015). Here we refer to Baraffe et al. (1998) for the evolution of the stellar luminosity. To calculate the profiles of heating rates and photolysis rates, we calculate the radiative transfer of parallel stellar photon beams in the spherically symmetric atmosphere by applying the method formulated by Tian et al. (2005a). Following this calculation, the spherical shell average is taken for the three-dimensional heating distribution. 

	We consider radiative cooling by thermal line emission of H$_{2}$O, OH, H$_{3}^{+}$, and OH$^{+}$ with line data provided by HITRAN database (Rothman et al., 2013; \url{http://hitran.org}) and ExoMol database (Tennyson et al., 2016; \url{http://exomol.com}). Our model contains 5021 transitions of H$_{2}$O, 383 transitions of OH, 5200 transitions of H$_{3}^{+}$, and 192 transitions of OH$^{+}$ to cover more than 99\% of the total energy emission in the temperature range of 100 - 1000 K. Then we calculate the radiative cooling rate by applying the method formulated by Yoshida \& Kuramoto (2020) which includes the Doppler shift of emission line wavelength caused by outflow acceleration. See Appendix A3 for the model formulation of radiative transfer processes. 
	
\begin{figure*}
\gridline{\fig{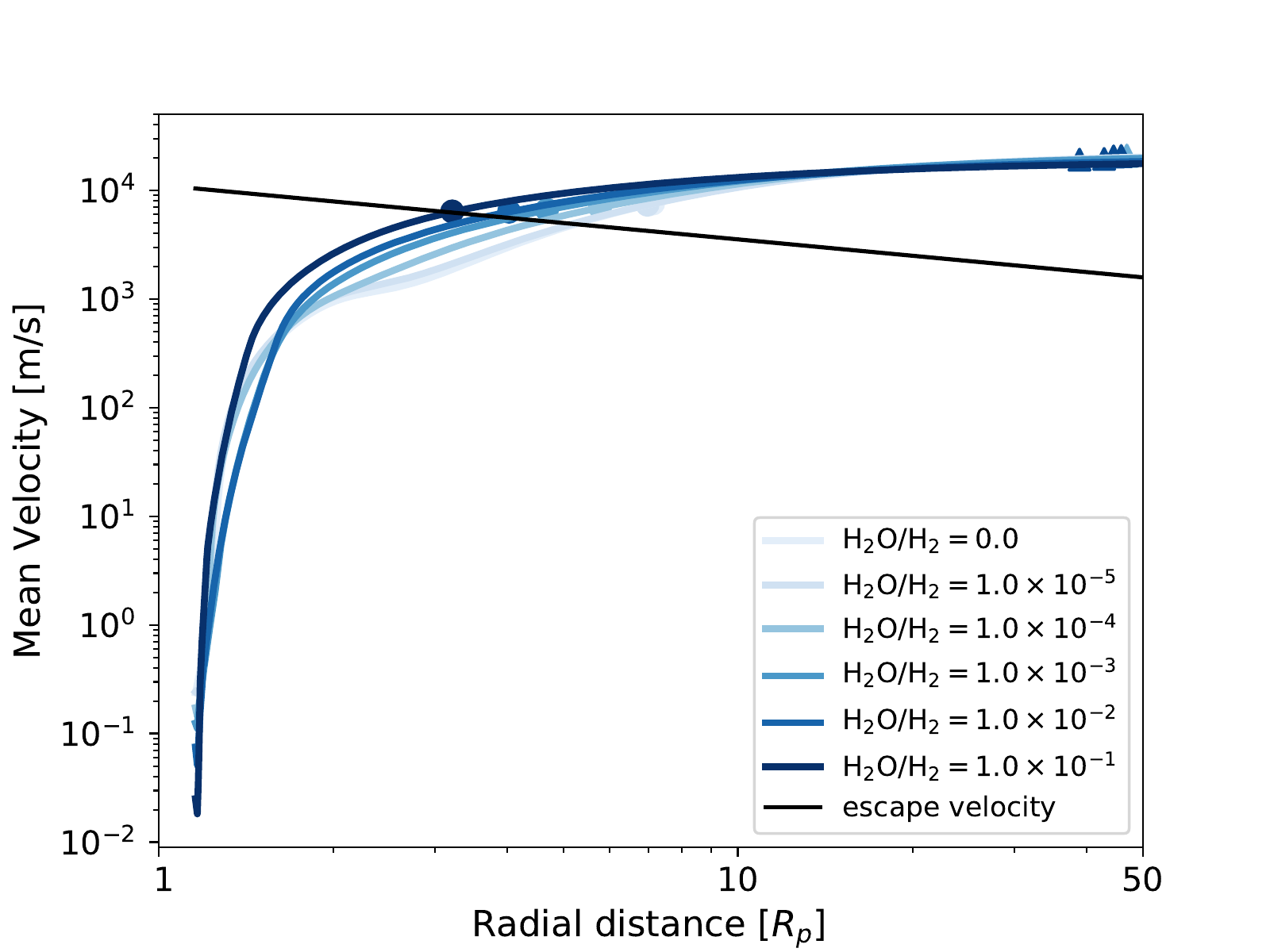}{0.45\textwidth}{(a)}
          \fig{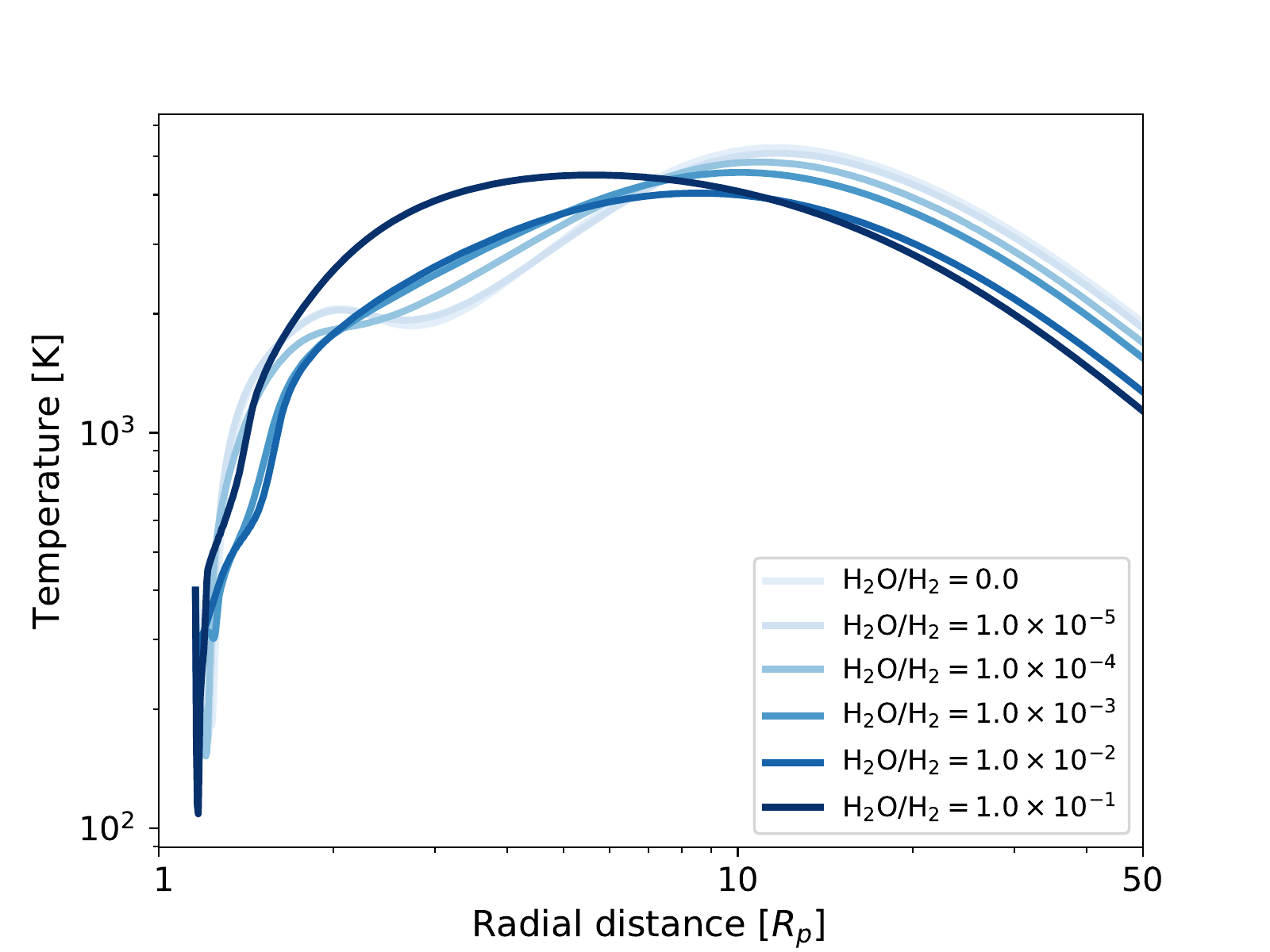}{0.45\textwidth}{(b)}
          }
\gridline{\fig{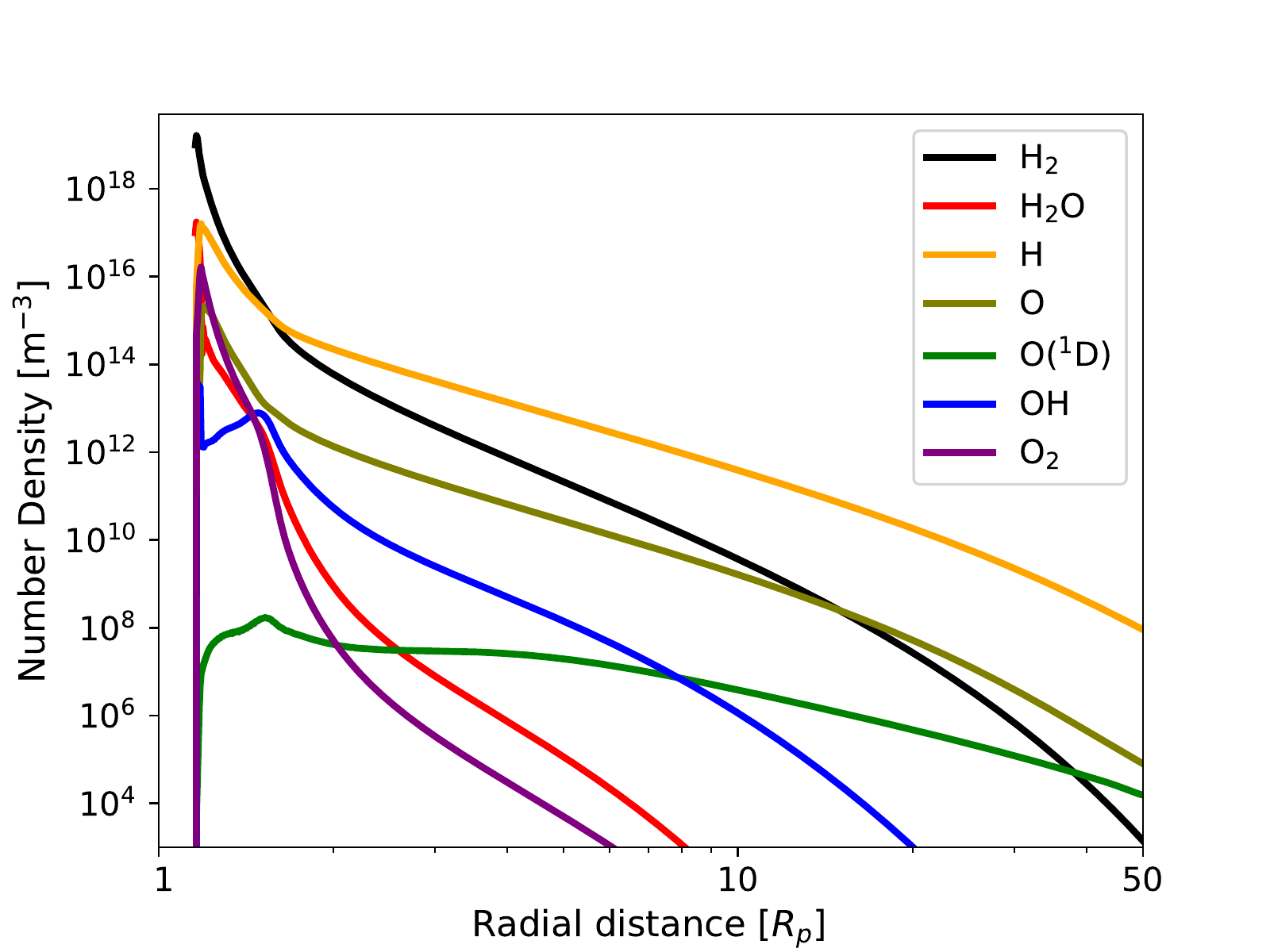}{0.45\textwidth}{(c)}
          \fig{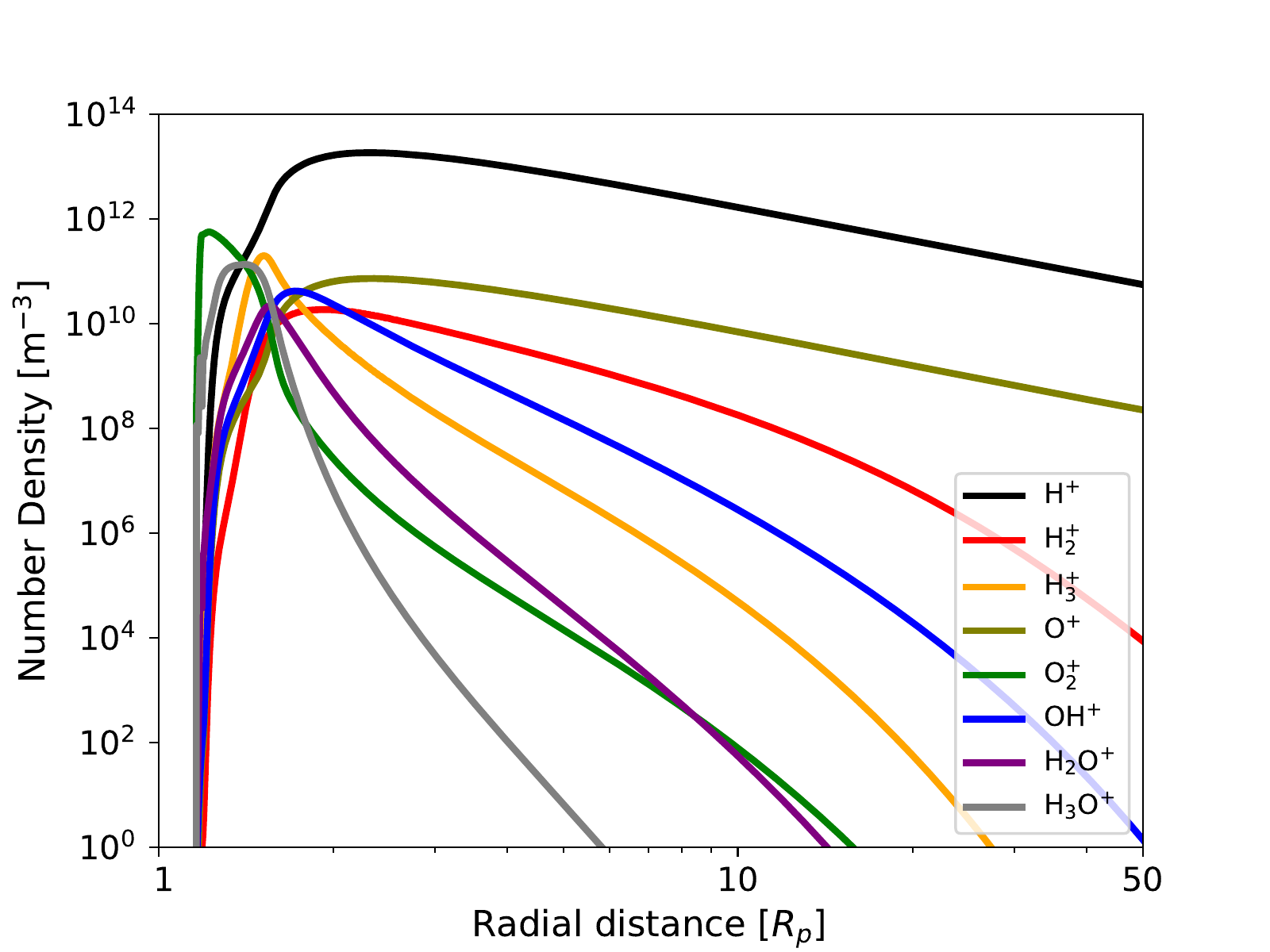}{0.45\textwidth}{(d)}
          }
\caption{Atmospheric profiles when the XUV flux is 100 times the present. (a) Mean velocity for six choices of the basal H$_{2}$O/H$_{2}$ mole ratio. The dots represent the transonic points. The triangles indicate the exobase. The black line represents the local escape velocity. (b) Temperature for six basal H$_{2}$O/H$_{2}$ ratios. (c) Number density of neutral species for H$_{2}$O/H$_{2}=0.01$. (d) Number density of ion species for H$_{2}$O/H$_{2}=0.01$. The radial distance is shown in the planet radius $R_{p}$.}
\label{fig:pyramid}
\end{figure*}

\section{Results} \label{sec:results}
\subsection{Atmospheric profiles and energy balance}
	The structure of steady outflows is shown in Fig. 1, which exhibits the radial profiles of mean velocity, temperature, and number densities of neutral and ion species for typical steady state solutions when the XUV flux is 100 times the present. The flow is radially accelerated from near zero velocity to supersonic in each solution (Fig. 1(a)). Molecules such as H$_{2}$ and H$_{2}$O are dissociated efficiently, and chemical products such as H, O, OH, and O$_{2}$ are produced (Fig. 1(c)). This behavior is common for other basal H$_{2}$O/H$_{2}$ cases. 
	
	The energy balance and its dependence on the basal H$_{2}$O/H$_{2}$ ratio in the lower region where the stellar XUV irradiation is absorbed mainly are shown in Fig. 2. The heating efficiency, $\eta$, which represents the net energy deposited as sensible heat relative to the total input of radiative energy, is given by
	\begin{equation}
		\eta =\frac{\int_{r_{0}}^{r_{1}}(q_{\mathrm{abs}}-q_{\mathrm{ch}}-q_{\mathrm{rad}})4\pi r^{2}dr}{\int_{r_{0}}^{r_{1}}q_{\mathrm{abs}}4\pi r^{2}dr},
	\end{equation}
	where $r$ is the radial distance from the center of the planet, $r_{0}$ is the radial distance of the lower boundary, $r_{1}=2R_{p}$ ($R_{p}$: planet radius), and $q_{\rm abs}$, $q_{\rm ch}$ and $q_{\rm rad}$ are the radiative heating rate by UV absorption, the rate of net chemical expense of energy, and the radiative cooling rate, respectively (see Appendix). As shown in the upper panel, the total radiative cooling rate increases with the basal H$_{2}$O/H$_{2}$ ratio due to the increase in the abundance of coolant. Here the main coolant is H$_{2}$O. In addition, radiatively active chemical products such as OH and OH$^{+}$ enhance the energy loss via radiative cooling. As a result, the heating efficiency decreases as the basal H$_{2}$O/H$_{2}$ ratio increases: it becomes lower than 10 \% when the basal H$_{2}$O/H$_{2}$ ratio $\geqq 0.1$ (see the lower panel in Fig. 2). 

\begin{figure}[htbp]
	\centering
	\includegraphics[width=0.45\columnwidth]{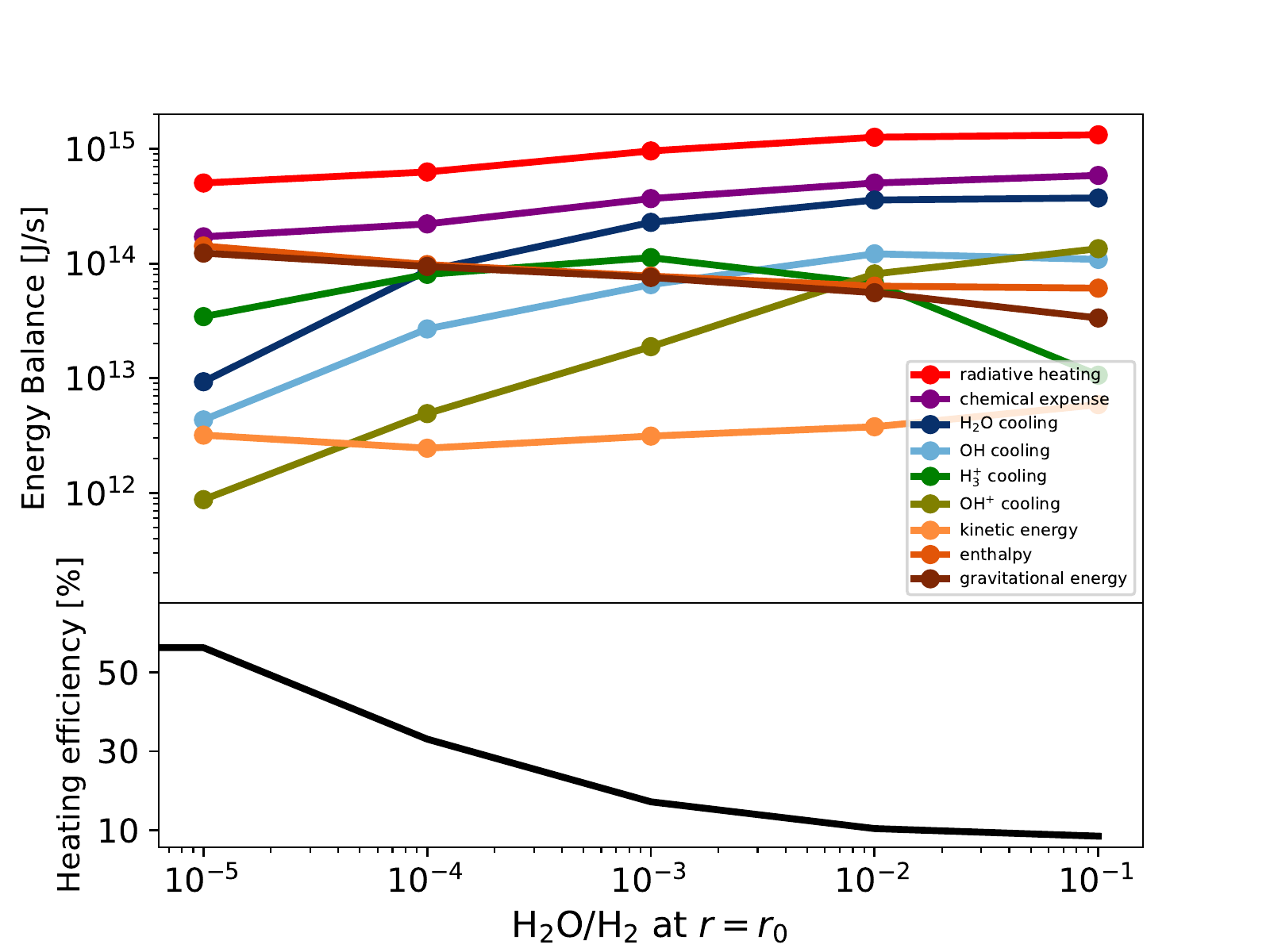}
	\caption{Energy balance in the region $r_{0}\leqq r \leqq 2R_{p}$ for different values of the basal H$_{2}$O/H$_{2}$ mole ratio when the XUV flux is 100 times the present; $R_{p}$ is the planet radius and $r_{0}$ is the radial distance of the atmospheric lower boundary ($=R_{p}+1000\,\mathrm{km}$). Legends: ``kinetic energy" is the flux of kinetic energy, ``enthalpy" is the flux of enthalpy, ``gravitational energy" is the difference of the gravitational potential, ``radiative heating" is the heating rate by UV absorption, ``chemical expense" is the net chemical expense of energy, and ``H$_{2}$O cooling", ``OH cooling", ``H$_{3}^{+}$ cooling" and ``OH$^{+}$ cooling" are the radiative cooling rate by H$_{2}$O, OH, H$_{3}^{+}$ and OH$^{+}$ respectively. The black line in the lower panel represents the heating efficiency (see Eq. (1) for its definition).}
\end{figure}

\subsection{Escape rate}
	The atmospheric escape rate in kg/Myr is shown in Fig. 3. Here the escape rate of species $i$ ($i=$H$_{2}$ or H$_{2}$O) is calculated by
	\begin{equation}
		F_{m,i}=4\pi r_{0}^{2}\rho_{i}(r_{0})v_{i}(r_{0}),
	\end{equation}
	where $\rho_{i}(r_{0})$ and $v_{i}(r_{0})$ are the mass density and velocity of species $i$ at the lower boundary, respectively. In the left panel, the escape rate of H$_{2}$ is found to decrease as the basal H$_{2}$O/H$_{2}$ ratio increases, which is due to the decrease in the heating efficiency by the enhanced radiative cooling and the suppression of the H$_{2}$ radial velocity by H$_{2}$O drag in the outflow. It is almost one order of magnitude smaller when H$_{2}$O/H$_{2}=0.1$ than that for the pure hydrogen atmosphere.
	
	While H$_{2}$O also escapes to space with H$_{2}$, fractionation between H$_{2}$ and H$_{2}$O occurs significantly as the basal H$_{2}$O/H$_{2}$ ratio increases, as noticed from a comparison between Fig. 3(a) and (b). The dashed line in Fig. 3(a) represents the critical flux of H$_{2}$, beyond which H$_{2}$ can drag H$_{2}$O up to outside the planetary gravitational well, given by
	\begin{equation}
		F_{\mathrm{crit,H_{2}O}}=\frac{4\pi r_{0}^{2}b_{\mathrm{H_{2},H_{2}O}}gX_{\rm H_{2}}}{k_{\rm B}T}(m_{\rm H_{2}O}-m_{\rm H_{2}}),
	\end{equation}
	where $k_{\rm B}$ is the Boltzmann constant, $m_{\rm H_{2}O}$ is the molecular mass of H$_{2}$O, $b_{\mathrm{H_{2},H_{2}O}}$ is the binary diffusion coefficient between H$_{2}$ and H$_{2}$O, $g$ is the gravity acceleration, and $X_{\rm H_{2}}$ is the mixing fraction of H$_{2}$ (Hunten et al., 1987). The escape rate of H$_{2}$ is found to become close to the critical flux as the basal H$_{2}$O/H$_{2}$ ratio increases. Our numerical integration becomes unstable when the escape rate of H$_{2}$ becomes close to the critical flux. The escape rate of H$_{2}$ would continue to decrease and reach the critical flux when the H$_{2}$O/H$_{2}$ ratio becomes enough large. Under larger H$_2$O/H$_2$ ratio, no escape of H$_{2}$O would occur. 
	
	The total escape rate is also found to increase almost proportionally to the XUV flux when the basal H$_{2}$O/H$_{2}$ ratio is the same. This is because the atmospheric profile and energy balance do not change significantly with the XUV flux. 
	
\begin{figure*}
\gridline{\fig{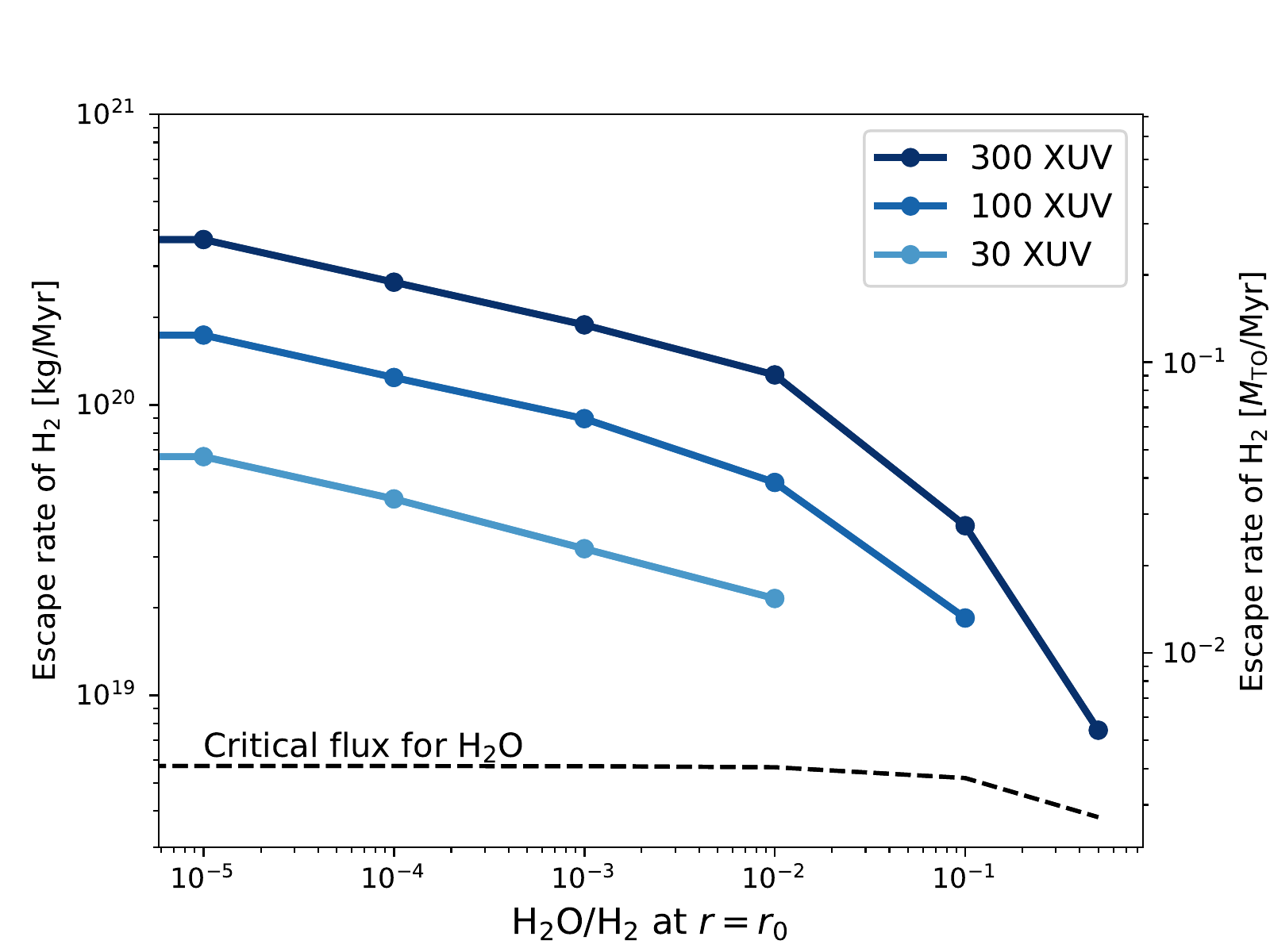}{0.45\textwidth}{(a)}
          \fig{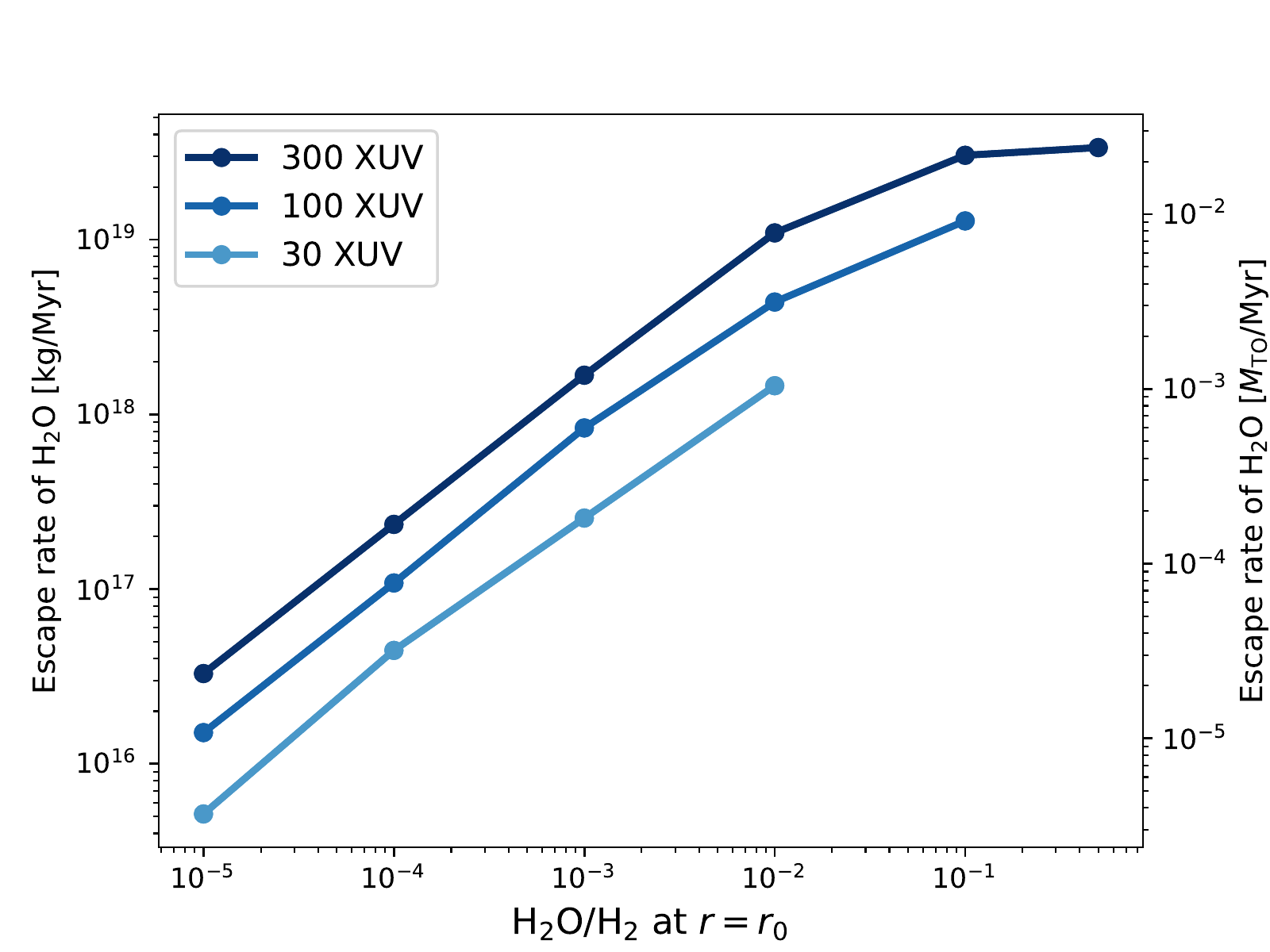}{0.45\textwidth}{(b)}
          }
\caption{Escape rates of H$_{2}$ (a) and H$_{2}$O (b) in kg/Myr as functions of the basal H$_{2}$O/H$_{2}$ mole ratio for the assumed XUV fluxes 30 times (``30 XUV"), 100 times (``100 XUV"), and 300 times (``300 XUV") the present. The right vertical axis represents the flux normalized by the mass of the present terrestrial ocean: $M_{\rm TO}=1.4\times 10^{21}\,\mathrm{kg}$. The dashed line in (a) represents the critical flux of H$_{2}$ above which H$_{2}$ can drag H$_{2}$O upward.
\label{fig:pyramid}}
\end{figure*}

\section{Discussion}
\subsection{Comparison with the results under the settings of radiative cooling processes used by Johnstone (2020)}
	In this section, we compare the results shown in Section 3 with those calculated under the settings of the radiative cooling processes used by Johnstone (2020) to clarify the effects of the radiative cooling by H$_{2}$O in vibrational bands and the radiatively active chemical products that we have considered newly. Johnstone (2020) considered the radiative emission by H$_{2}$O in rotational bands formulated by Hollenbach and McKee (1979), Ly-$\alpha$ emission by H given by Murray-Clay et al. (2009) and Guo (2019), and emission by O at 63 and 147 $\mathrm{\mu m}$ derived by Bates (1951). Figure 4 shows the difference in the energy balance and atmospheric escape rate between the results under the settings of radiative emission of this study and those under Johnstone's settings. The radiative cooling rate of H$_{2}$O under our settings is larger than that under Johnstone's settings due to the addition of the emission in vibrational bands. Moreover, the radiative emission by the chemical products that we have considered enhances the radiative cooling rate, while the emission by H and O in Johnstone's settings has little effect on the energy balance. As a result, the heating efficiency and atmospheric escape rate under our settings become lower over the range of the atmospheric compositions that we have considered. 
	
\begin{figure*}
\gridline{\fig{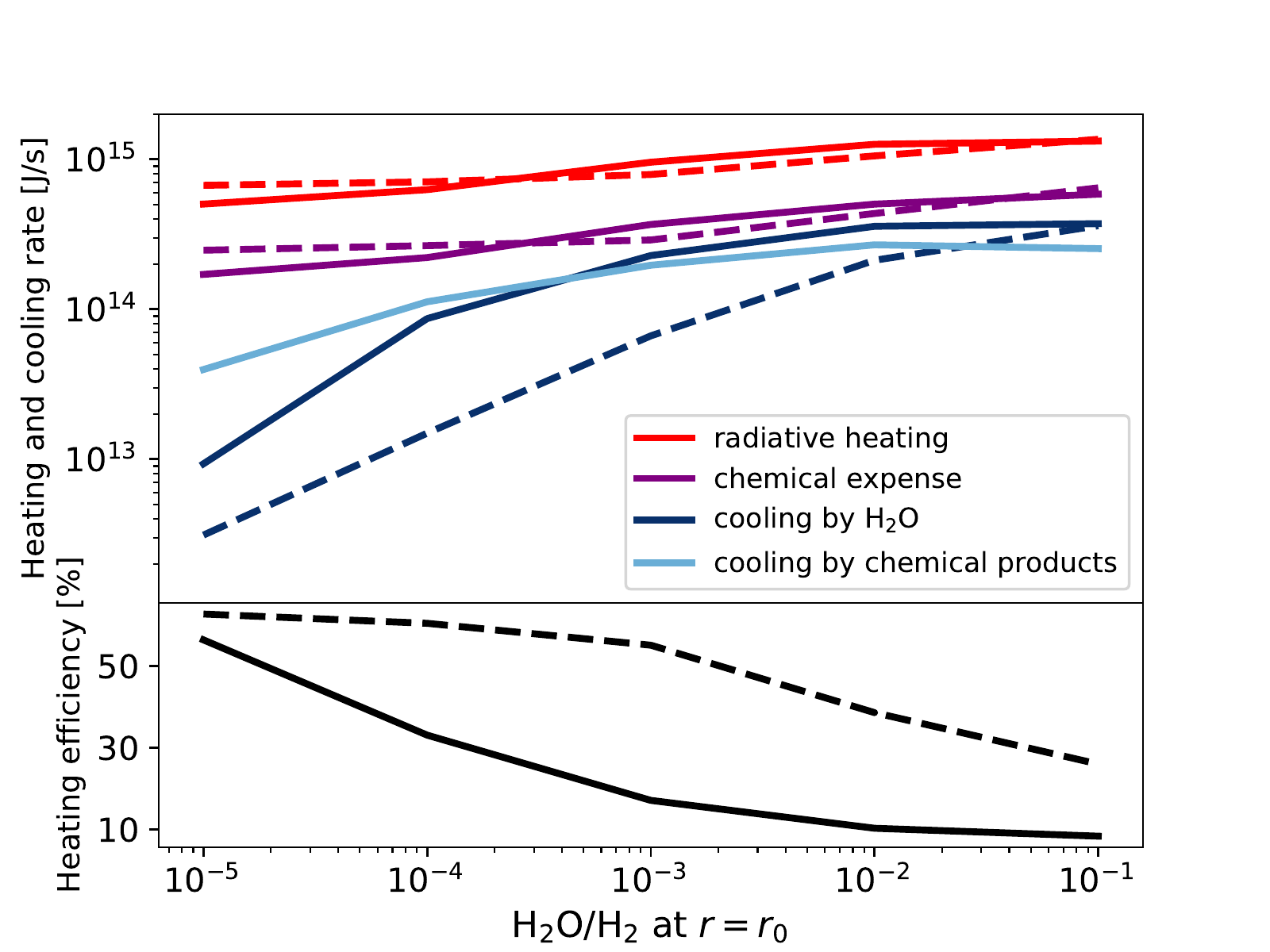}{0.45\textwidth}{(a)}
          \fig{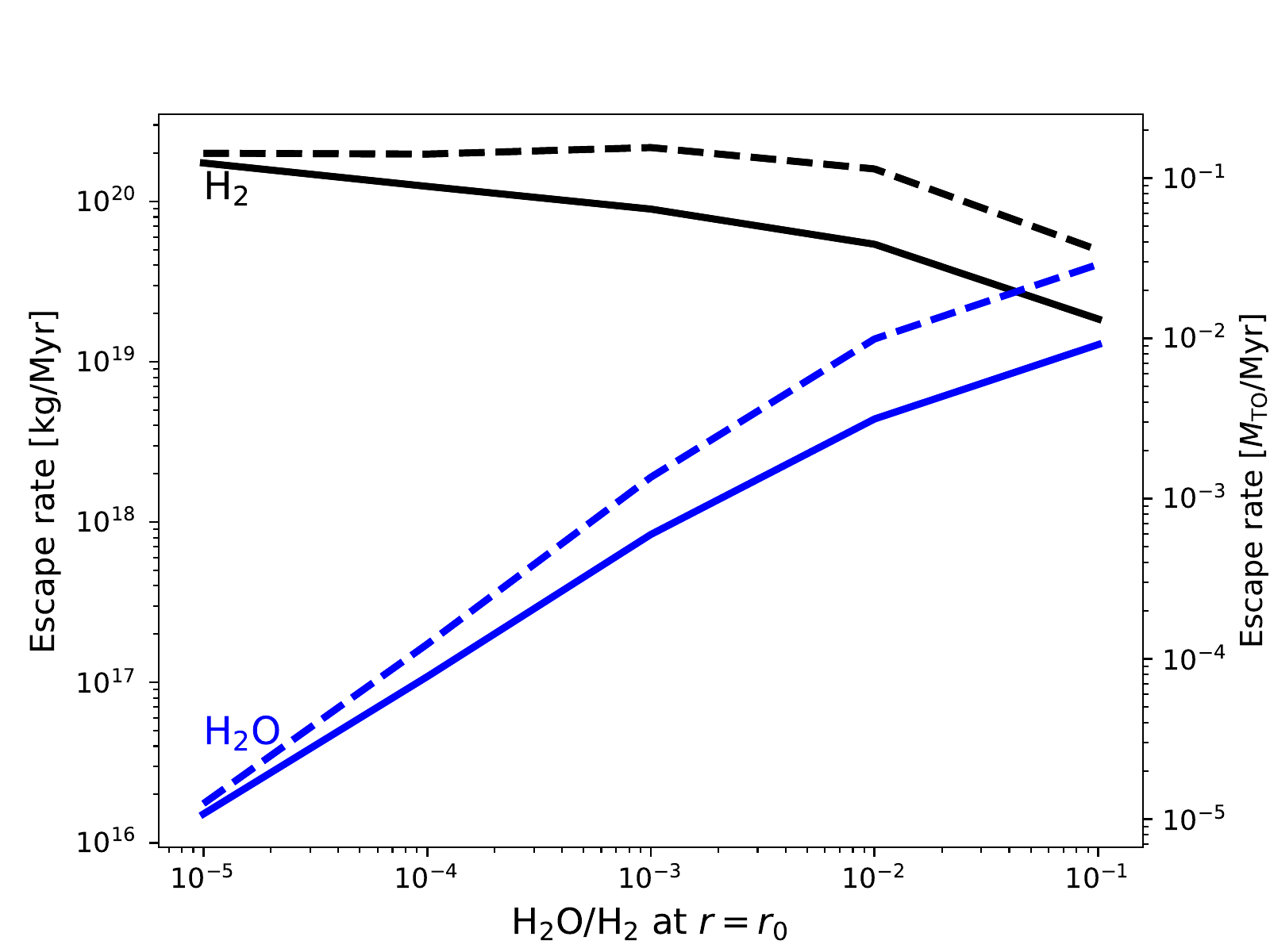}{0.45\textwidth}{(b)}
          }
\caption{Radiative heating and cooling rates (upper panel) and heating efficiencies (lower panel) in the region $r_{0}\leqq r \leqq 2R_{p}$ (a) and atmospheric escape rates in kg/Myr (b) as functions of the basal H$_{2}$O/H$_{2}$ mole ratio when the XUV flux is 100 times the present. The right vertical axis of (b) represents the flux normalized by the mass of the present terrestrial ocean: $M_{\rm TO}=1.4\times 10^{21}\,\mathrm{kg}$. The solid lines correspond to the values under the settings of radiative emission of this study, and the dashed lines correspond to those under the settings of Johnstone (2020).
\label{fig:pyramid}}
\end{figure*}

\subsection{Evolution of H$_{2}$-H$_{2}$O atmospheres during the pre-main sequence phase}
\subsubsection{Changes in the amount and composition of H$_{2}$-H$_{2}$O atmospheres}
	In this section, we show the evolution of H$_{2}$-H$_{2}$O atmospheres in their runaway greenhouse condition during the pre-main sequence phase of the host star based on our new estimate of the atmospheric escape rate. Firstly, we consider a planet with the mass and radius same as those of the Earth, orbiting at 0.02 au around a pre-main sequence star that ends up an M8 dwarf. This orbit corresponds to the inner edge of the habitable zone after the star entered the main sequence phase, so the planet is exposed to extremely strong XUV flux during the pre-main sequence phase and moreover the duration of the runaway phase is quite long. We assume that the duration of the runaway greenhouse phase is 2 Gyr (Ramirez and Kaltenegger, 2014). The planet is assumed to be exposed to the stellar irradiation since the stellar age reaches 10 Myr when the surrounding nebular gas is expected to have dissipated (e.g., Haisch et al., 2001). The XUV flux during the pre-main sequence phase is assumed to be 0.001 times the total stellar luminosity: it decreases from 230 times to 10 times the present as the stellar age increases from 10 Myr to 1 Gyr (Fig. 1 in Luger and Barnes, 2015). The lower atmosphere is assumed to be composed of H$_{2}$ and H$_{2}$O. The mixing ratio H$_{2}$O/H$_{2}$ and the initial atmospheric amount are taken to be parameters because of the uncertainty in the origin and supply of volatiles. For simplicity, we neglect other atmospheric escape processes, volatile delivery, and interaction between the atmosphere and the planetary surface during the hydrodynamic escape. The composition at the lower boundary of the upper escaping atmosphere is assumed to be equal to the composition at the surface. Once the H$_{2}$ escape flux reaches the critical flux for H$_{2}$O (see Eq. 3), we stop the H$_{2}$O escape and consider diffusion-limited escape for H$_{2}$. As mentioned, however, we cannot find the solution with the H2 escape flux equal to the critical flux, because of numerical instability; instead, we extrapolate the highest value of the H$_{2}$O/H$_{2}$ ratio at which we could find the solution toward higher values (namely, extending the solid lines to find the crossover point with the dashed line in Fig. 3).
	
	Examples of atmospheric evolutionary tracks are shown in Fig. 5. The atmospheric mass of each species is expressed by the mass of the present terrestrial ocean: $M_{\rm TO}=1.4\times 10^{21}\,\mathrm{kg}$. The amount of H$_{2}$ decreases with time gradually. On the other hand, the decrease in the amount of H$_{2}$O stops, when the H$_{2}$ escape flux reaches the critical flux for H$_{2}$O. In some cases, both H$_{2}$ and H$_{2}$O are left behind when the runaway greenhouse phase ends (=2000 Myr), indicating that an ocean is formed and an H$_{2}$-rich atmosphere remains after the star entered the main sequence phase. 		
	
	The timescale for H$_{2}$ escape and the final H$_{2}$O amount are shown as functions of the initial atmospheric amounts of H$_{2}$ and H$_{2}$O in Fig. 6. Here we define the timescale for H$_{2}$ escape as the time for the amount of H$_{2}$ to reach $1.0\times 10^{-3}\,M_{\rm TO}$. As seen in Fig. 6(a), the timescale for H$_{2}$ escape becomes longer as the initial H$_{2}$O/H$_{2}$ ratio increases. This is due to the suppression of the atmospheric escape by the radiative cooling and the decrease in the diffusion velocity of H$_{2}$. The final H$_{2}$O amount also increases as the initial H$_{2}$O/H$_{2}$ ratio increases (Fig. 6(b)). The timescale for H$_{2}$ escape exceeds the duration of the runaway greenhouse phase and, thus, part of H$_{2}$O is left behind in a wide range of initial atmospheric amounts.
	
	The formation of such massive H$_{2}$-H$_{2}$O atmospheres that we have considered here could occur depending on planetary accretion processes, nebular properties, and so on. If a proto-planet reaches a mass of $\sim 1M_{\oplus}$ before the protoplanetary nebula dissipates, it can capture the surrounding nebular gas with more than $\sim 10M_{\rm TO}$ (e.g., Ikoma and Genda, 2006). Moreover, the mass of the nebula-captured atmosphere may increase significantly when the atmospheric mean molecular weight and effective specific heat increase by mixing of high-molecular-weight components such as H$_{2}$O (Kimura and Ikoma, 2020). In addition, volatile-rich planetary building blocks like carbonaceous chondrites could also deliver volatiles although their amounts are highly uncertain.
	
	Figure 7 is the same as Fig. 6 but for the planetary orbital radius of 0.05 au, which corresponds to the outer edge of the habitable zone around M8 dwarfs (Luger and Barnes, 2015). In this case, H$_{2}$ and H$_{2}$O tend to be left behind after the runaway greenhouse phase compared with the planet in the inner edge of the habitable zone because the incident XUV flux is smaller and the duration of the runaway greenhouse condition is shorter ($\sim 0.1\,\mathrm{Gyr}$; Luger and Barnes, 2015). Planets around more massive M dwarfs are also likely easy to keep H$_{2}$ and H$_{2}$O until the runaway greenhouse condition ends for the same reason. 
	
\begin{figure}
	\centering
	\includegraphics[width=0.45\columnwidth]{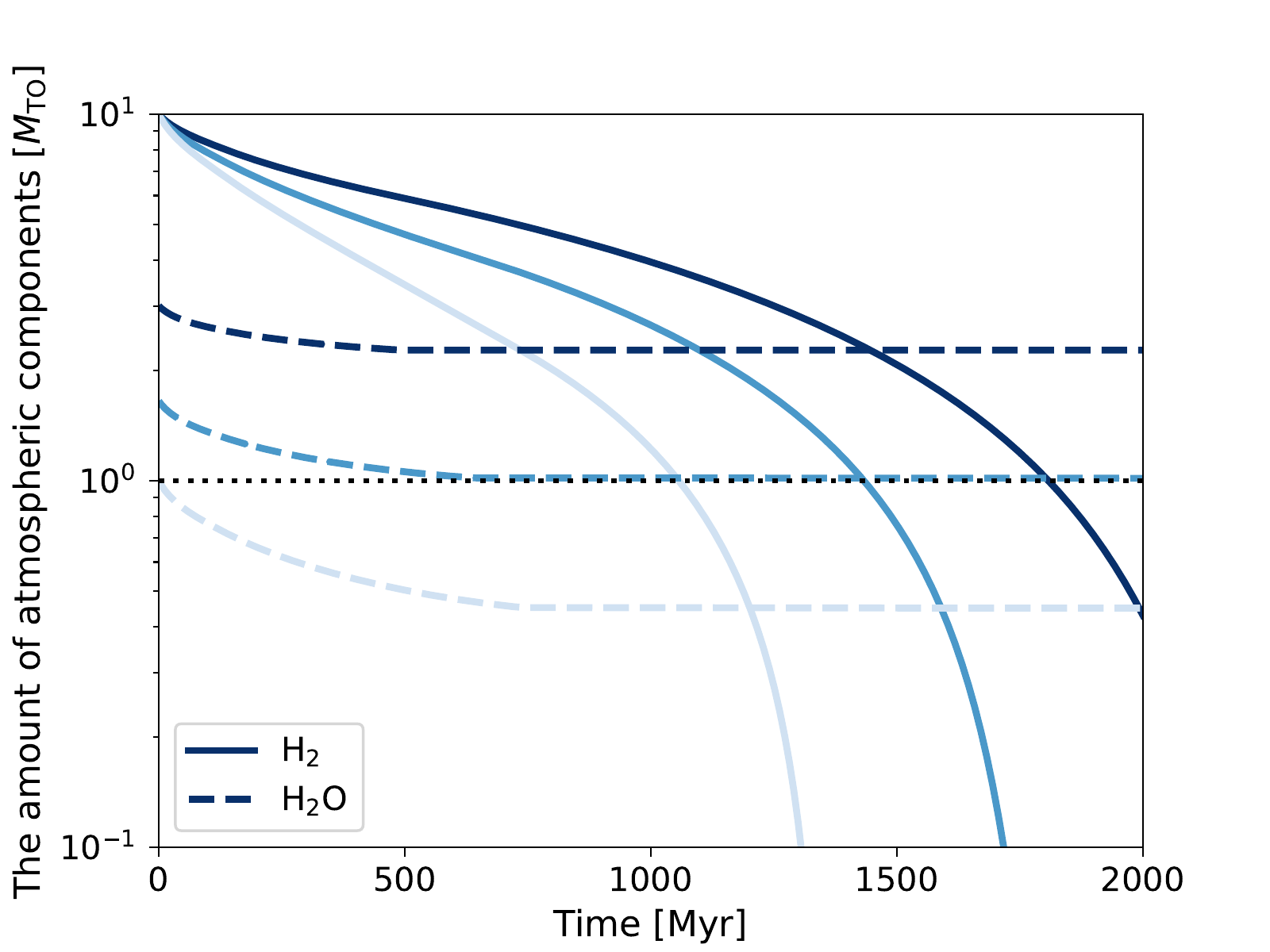}
	\caption{Changes in the atmospheric amounts of H$_{2}$ and H$_{2}$O with time. The vertical axis represents the atmospheric mass of each species normalized by the mass of the present terrestrial ocean: $M_{\rm TO}=1.4\times 10^{21}\,\mathrm{kg}$. The solid lines and dashed lines represent the amounts of H$_{2}$ and H$_{2}$O, respectively. The initial amount of H$_{2}$ is $10M_{\rm TO}$, and the initial amounts of H$_{2}$O are $1M_{\rm TO},\,1.7M_{\rm TO},\,3M_{\rm TO}$, respectively.}
\end{figure}

\begin{figure*}
\gridline{\fig{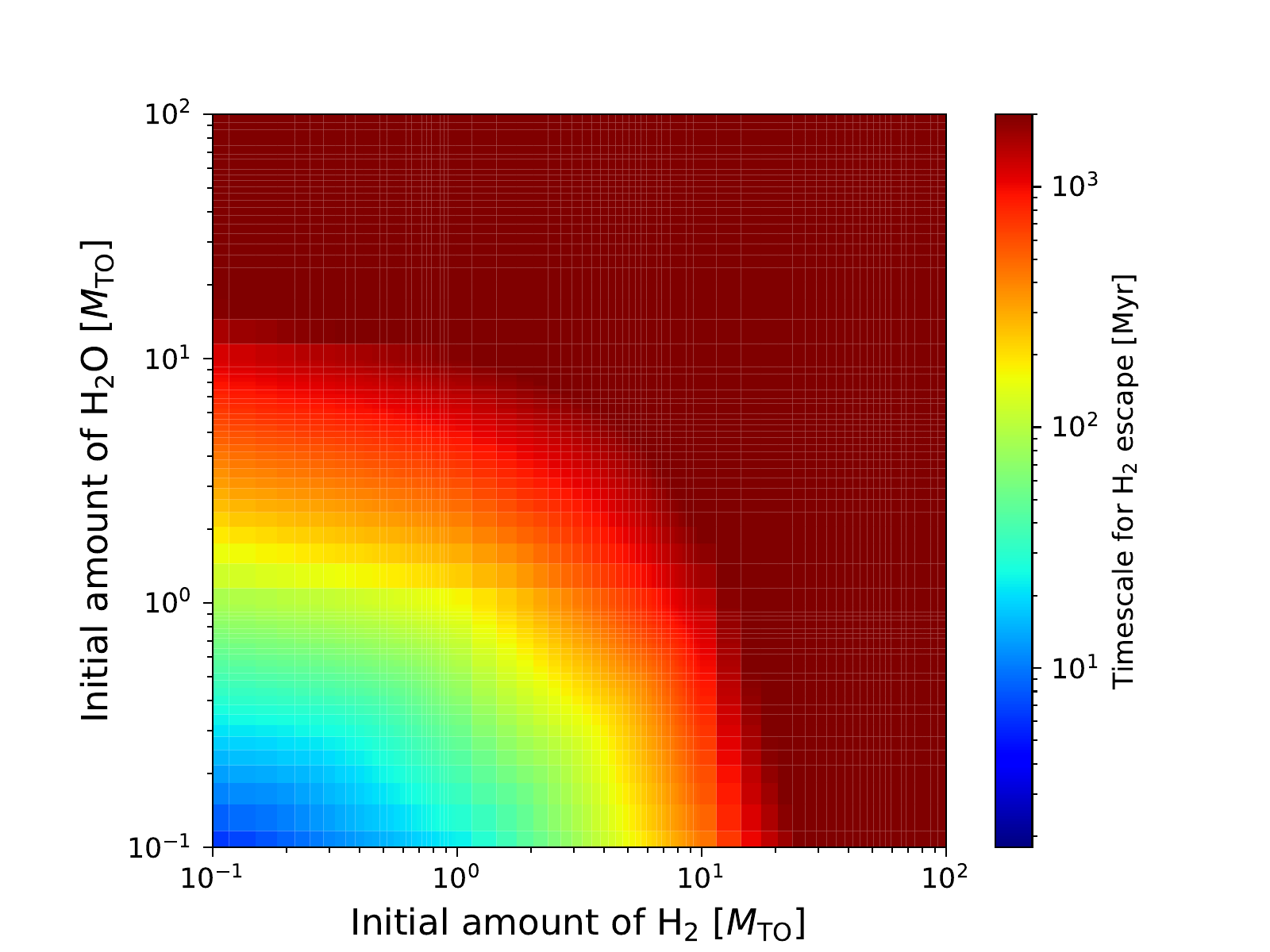}{0.45\textwidth}{(a)}
          \fig{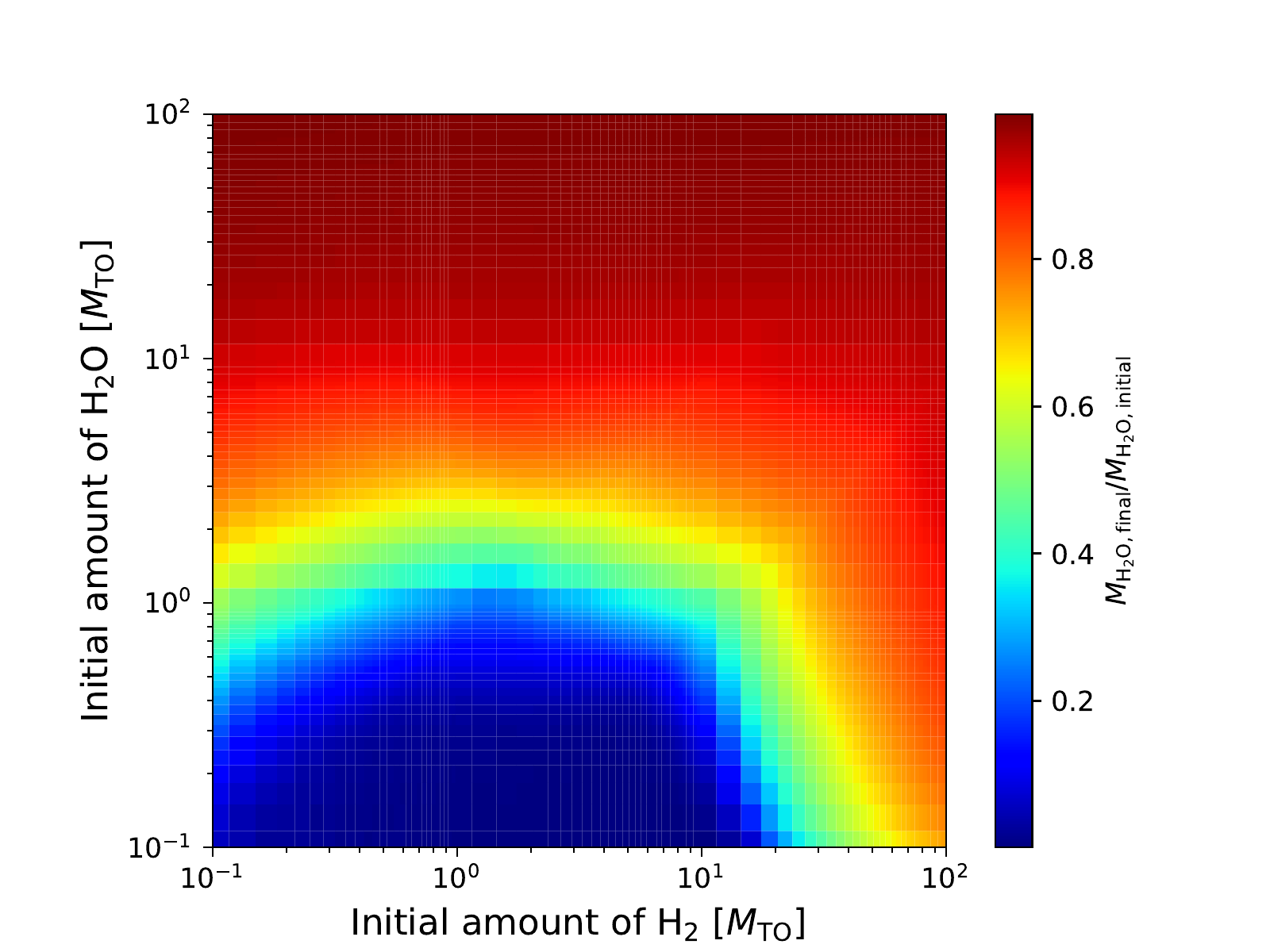}{0.45\textwidth}{(b)}
          }
\caption{Timescale for H$_{2}$ escape (a) and the final H$_{2}$O amount relative to its initial amount (b) as functions of the initial amounts of H$_{2}$ and H$_{2}$O on the planet of $1 M_{\oplus}$ orbiting at 0.02 au. The duration of the runaway greenhouse phase is assumed to be 2 Gyr.
\label{fig:pyramid}}
\end{figure*}

\begin{figure*}
\gridline{\fig{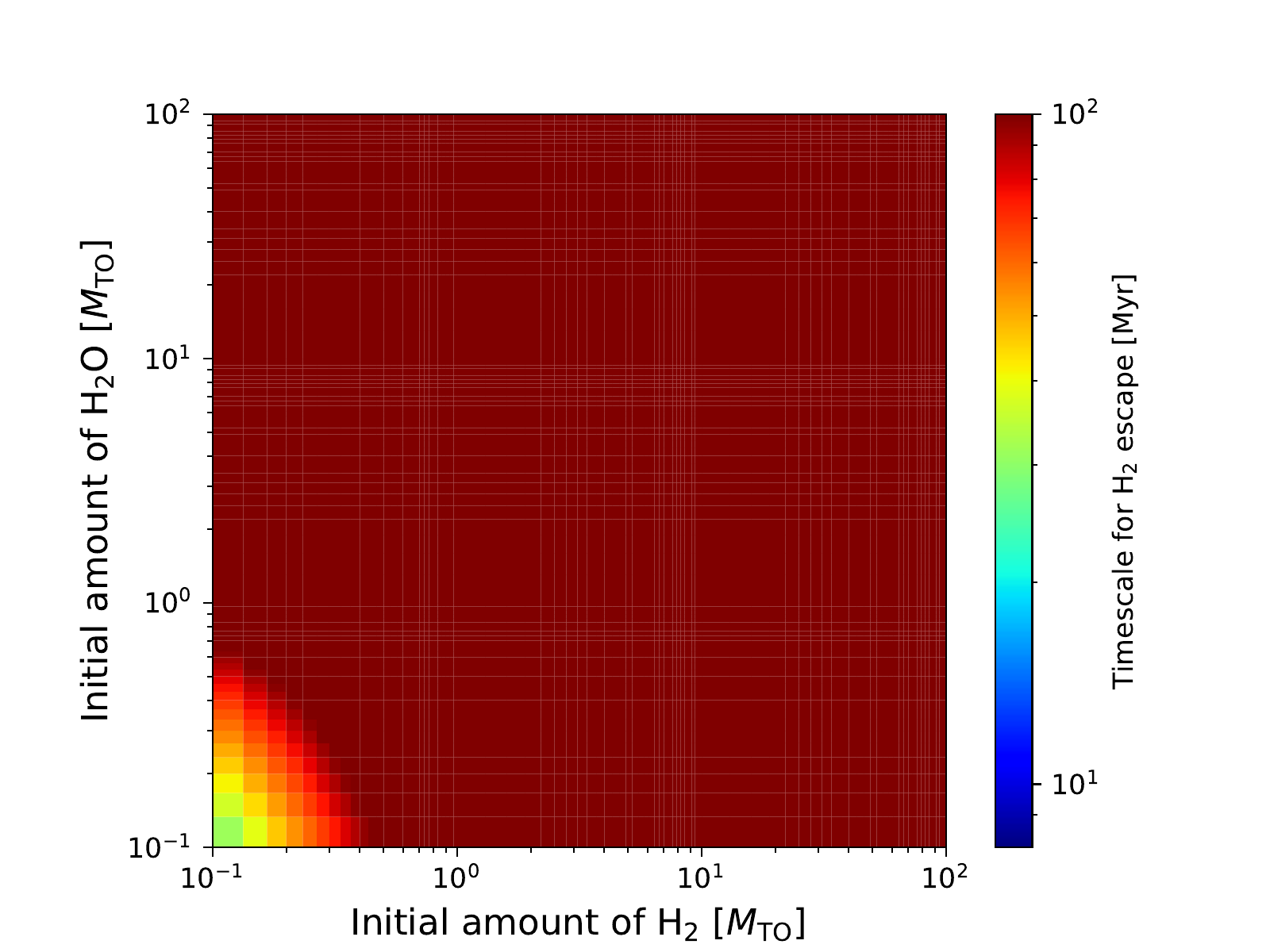}{0.45\textwidth}{(a)}
          \fig{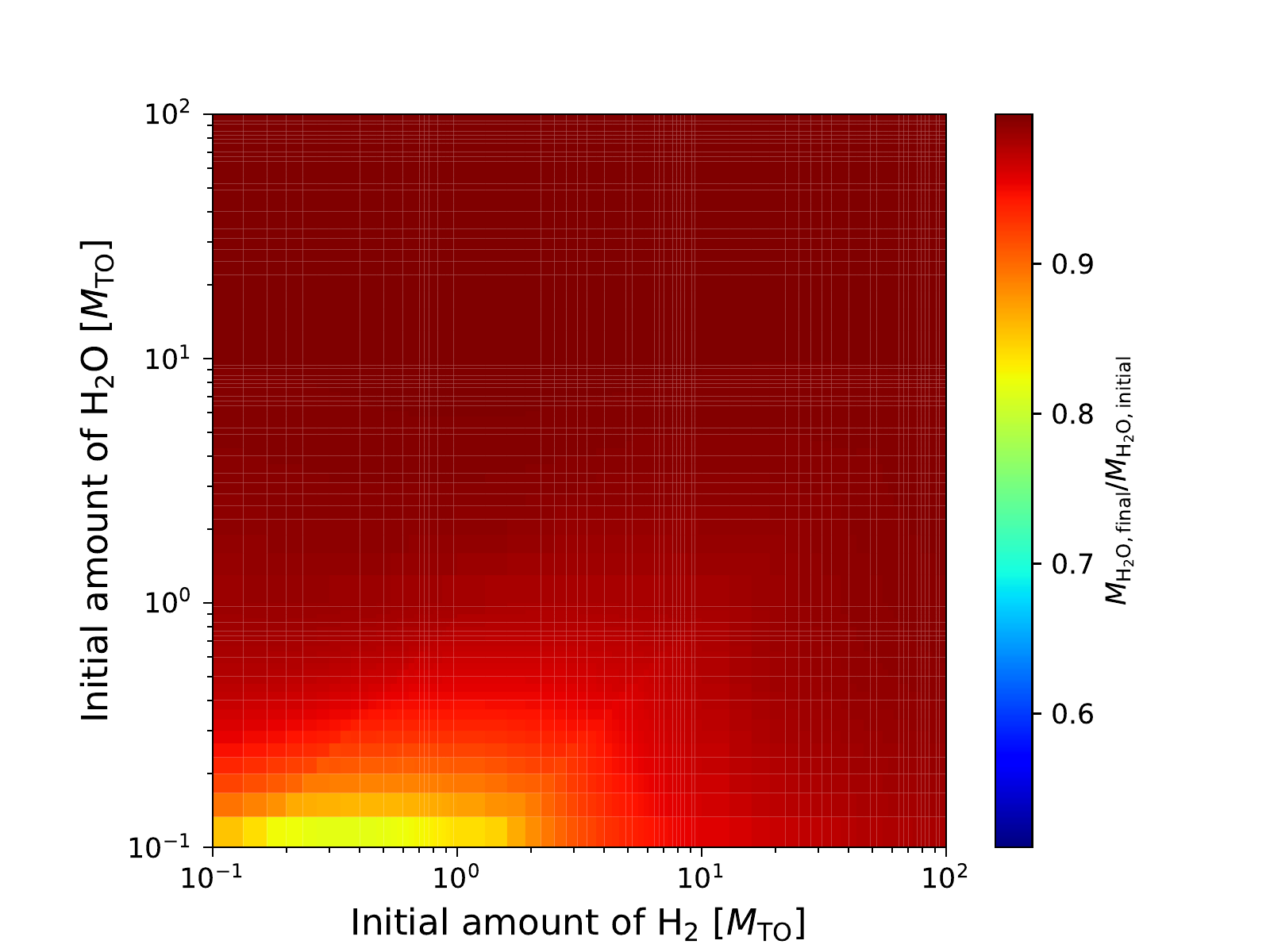}{0.45\textwidth}{(b)}
          }
\caption{Same as Fig. 6 but for the planet orbiting at 0.05 au. The duration of the runaway greenhouse phase is assumed to be 0.1 Gyr.
\label{fig:pyramid}}
\end{figure*}	
	
\subsubsection{Other processes affecting the estimation of atmospheric evolution}
	In this study, we consider only H$_{2}$, H$_{2}$O, and their chemical products as the atmospheric components for simplicity. Qualitatively, considering other components such as carbon species leads to a decrease in the atmospheric escape rate by their strong radiative cooling (Yoshida and Kuramoto, 2020; 2021). Moreover, radiatively active ions such as H$^{-}$ and H$_{3}$O$^{+}$ could enhance the effect of the radiative cooling (e.g., Lenzuni et al., 1991; Yurchenko et al., 2020). Quantitative investigation of their effects is beyond the scope of this study.
	
	We neglect interaction between the atmosphere and the planetary surface. Planets with runaway greenhouse atmospheres are predicted to have magma oceans (e.g., Hamano et al., 2013; Lebrun et al., 2013). In that case, if material circulation and mixing occur efficiently both in the atmosphere and magma ocean, the atmospheric composition is buffered by the redox state of the mantle (e.g., Holland, 1984). It is our future work to consider the effects of the interaction between the atmosphere and surface on the atmospheric evolution.
	
	We assume that the atmospheric composition at the lower boundary in the calculations is equal to the composition at the surface. However, chemical processes in the lower and middle atmosphere could change the composition at the bottom of the escaping upper atmosphere from that at the surface. Part of H$_{2}$O is expected to be dissociated by photolysis in the stratosphere (e.g., Hunten and Strobel, 1974), which leads to decreasing the mixing ratio of H$_{2}$O in the escape outflow, and increasing the efficiency of H$_{2}$ escape. 
	
	Provided the H$_{2}$O-dominated atmosphere is kept in a moist/runaway-greenhouse condition, the remaining H$_{2}$O would continue to be lost through its photolysis followed by escape of hydrogen produced from H$_{2}$O. This process may have also been important for water loss from Venus. Our model estimates that H$_{2}$ is lost almost completely in $\sim 1$ Gyr from an H$_{2}$-H$_{2}$O atmosphere, if the planet has been located at the current location of Venus from a Sun-like young star and the initial atmospheric mass is a few times as massive as the present-day Earth’s oceans. Given that Venus is expected to have been in a moist/runaway-greenhouse condition, it can lose H$_{2}$O completely in the remaining $\sim 3$ Gyr (e.g., Kasting et al., 1993; Hamano et al., 2013). This is a clear contrast to terrestrial planets currently orbiting in the habitable zones around M dwarfs, which ended the moist/runaway-greenhouse condition, leaving H$_{2}$O due to cold trap at the time when their host stars entered the main sequence phase. According to classic models (e.g., Hunten, 1973; Kasting et al., 1993), the timescale for the complete loss of H$_{2}$O with initial amount equivalent to that of the present-day Earth’s oceans is several hundred million years, provided H$_{2}$O is a major component in the atmosphere and hydrogen produced from H$_{2}$O escapes to space in a diffusion-limited fashion. On the other hand, a large amount of oxygen produced from photolysis of H$_{2}$O may be left and accumulated in the remaining atmosphere, leading to a significant reduction in the diffusion velocity and, thus, the escape rate of hydrogen. Further investigation of the escape process of the H$_{2}$O-dominated atmosphere is needed.
	
\subsubsection{Implication for habitability of planets around M dwarfs}
	Previous studies for the evolution of pure H$_{2}$O atmospheres of Earth-mass rocky planets orbiting pre-main sequence M dwarfs showed that several times the amount of Earth's seawater is lost and massive O$_{2}$ could be produced photochemically (e.g., Luger and Barnes, 2015). The planetary desiccation could severely hamper the ability of life to originate and evolve on planets. Moreover, oxidizing environments could also prevent the emergence of life organisms because prebiotic chemistry needs reducing environments (e.g., Schlesinger and Miller, 1983). Therefore, their results indicated that terrestrial planets around M dwarfs are not suitable for the origin and evolution of life.

	By contrast, our results indicate that both H$_{2}$ and H$_{2}$O can survive during the early runaway greenhouse phase and H$_{2}$-rich environments with oceans could be formed after the end of the runaway greenhouse phase. Such reducing environments are suitable for photochemical production of organic matters potentially linked to the emergence of living organisms (e.g., Schlesinger and Miller, 1983). Therefore, some planets in the habitable zones around M dwarfs may have temperate and reducing surface environments suitable for the origin of life.

\section{Conclusion}
	We have applied our 1-D hydrodynamic escape model to an H$_{2}$-H$_{2}$O atmosphere on a planet with mass of $1M_{\oplus}$ around a pre-main sequence M dwarf. According to our results, the atmospheric escape rate decreases with the basal H$_{2}$O/H$_{2}$ ratio because of the energy loss by the radiative cooling of H$_{2}$O and chemical products such as OH and H$_{3}^{+}$. The escape rate of H$_{2}$ becomes one order of magnitude smaller when the basal H$_{2}$O/H$_{2}=0.1$ than that of the pure hydrogen atmosphere. The timescale for H$_{2}$ escape could exceed the duration of the early runaway phase depending on the initial atmospheric amount and composition. Our results suggest that temperate and reducing environments with oceans which are suitable for the origin of life could be formed after the end of the runaway greenhouse phase on terrestrial planets around M dwarfs.

\section*{Acknowledgements}
	We thank an anonymous reviewer whose comments greatly improved the manuscript. This work was supported by MEXT/JSPS KAKENHI Grant Number 18H05439. N. T. was also supported by JSPS KAKENHI Grant Number 18KK0093, 19H00707, 20H00192, and 22H00164. M. I. was also supported by JSPS KAKENHI Grant Number JP21H01141. K. K. was also supported by MEXT/JSPS KAKENHI Grant Number 17H06457 and 21K03638.

\appendix
\section{Details of the model}
\subsection{Basic equations}
	We have solved the equations of continuity, momentum and energy for a multi-component gas assuming spherical symmetry,
 	\begin{equation}
 		\frac{\partial n_{i}}{\partial t}+\frac{1}{r^{2}}\frac{\partial (n_{i} u_{i}r^{2})}{\partial r}=\omega_{i},
 	\end{equation}
 	\begin{equation}
 		\frac{\partial u_{i}}{\partial t}+u_{i}\frac{\partial u_{i}}{\partial r}=-\frac{1}{\rho_{i}}\frac{\partial p_{i}}{\partial r}-\frac{GM}{r^{2}}+\sum_{\substack{j}}(u_{j}-u_{i})\frac{n_{j}}{m_{i}}\mu_{ij}k_{ij},
 	\end{equation}
 	\begin{equation}
 		\frac{\partial}{\partial t}\left[\rho \left(\frac{1}{2}u^{2}+E\right)\right]+\frac{1}{r^{2}}\frac{\partial}{\partial r}\left\{\left[\left(\frac{1}{2}u^{2}+E+\frac{p}{\rho}\right)\rho u \right]r^{2}\right\} =-\frac{GM}{r^{2}}\rho u+q,
 	\end{equation}
	where $t$ is the time, $r$ is the distance from the planet's center, $n_{i}, \rho_{i}, u_{i}, p_{i}$, and $\omega_{i}$ are the number density, mass density, velocity, partial pressure, and production rate of species $i$, respectively, $G$ is the gravitational constant, $M$ is the mass of the planet, $\rho$, $p$, and $E$ are the total mass density,  total pressure, and total specific internal energy, respectively, $u$ is the mean gas velocity, $\mu_{ij}$ is the reduced molecular mass between species $i$ and $j$, $k_{ij}$ is the momentum transfer collision frequency that follows
	\begin{equation}
 		k_{ij} = \frac{k_{\rm B}T}{\mu_{ij}b_{ij}},
 	\end{equation}
	where $k_{\mathrm{B}}$ is the Boltzmann constant, $T$ is the temperature and $b_{ij}$ is the binary diffusion coefficient, which is given by (Zahnle et al., 1990)
	\begin{equation}
  		b_{ij} = \begin{cases}
   			1.96\times 10^{6}\frac{T^{1/2}}{\mu_{ij}^{1/2}}\hspace{32.5pt}\mathrm{for\,neutral \mathchar`- neutral\,pair} \\
			4.13\times 10^{-8}\frac{T}{\mu_{ij}^{1/2}\alpha^{1/2}}\hspace{14pt}\mathrm{for\,neutral \mathchar`- ion\,pair} \\
			8.37\times 10^{-5}\frac{T^{5/2}}{\mu_{ij}^{1/2}}\hspace{27pt}\mathrm{for\,ion \mathchar`- ion\,pair}
  		\end{cases}
	\end{equation}
	where $\alpha$ is the polarizability (Banks and Kockarts, 1973; Garcia, 2007). Here, $b_{ij}$, $T$, $\mu_{ij}$ and $\alpha$ are expressed in $\mathrm{cm}^{-1}\mathrm{s}^{-1}$, $\mathrm{K}$, $\mathrm{g}$ and $\mathrm{cm^{3}}$, respectively. The total internal energy is given by	
	\begin{equation}
		\rho E= \frac{1}{n}\left(\sum_{\substack{i}}\frac{n_{i}}{\gamma_{i}-1}\right)p,
 	\end{equation}
	where $\gamma_{i}$ is the ratio of specific heat of species $i$. The net heating rate is given by 
	\begin{equation}
		q=q_{\mathrm{abs}}-q_{\mathrm{ch}}-q_{\mathrm{rad}},
 	\end{equation}
	where $q_{\mathrm{abs}}$ is the heating rate by XUV absorption (see A.3), $q_{\mathrm{ch}}$ is the rate of net chemical expense of energy (see A.2), and $q_{\mathrm{rad}}$ is the radiative cooling rate by infrared active molecules (see A.3).
	
\subsection{Chemical processes}
	A total of 93 chemical reactions are considered for 15 atmospheric components: H$_{2}$, H$_{2}$O, H, O, O($^{1}$D), OH, O$_{2}$, H$^{+}$, H$_{2}^{+}$, H$_{3}^{+}$, O$^{+}$, O$_{2}^{+}$, OH$^{+}$, H$_{2}$O$^{+}$, and H$_{3}$O$^{+}$ (Table A1). 
	
	The photolysis rate $f_{\mathrm{ph},i}$ is given by
	\begin{equation}
		f_{\mathrm{ph},i}=\int_{\lambda}n_{i}\sigma_{i}(\lambda)I(\lambda)d\lambda,
 	\end{equation}
	where $\sigma_{i}(\lambda)$ is the photodissociation/photoionization cross section at wavelength $\lambda$ for species $i$ and $I(\lambda)$ is the incident XUV photon flux at wavelength $\lambda$. The energy consumed per unit time by this photolysis $q_{\mathrm{ph}, i}$ is given by
	\begin{equation}
		q_{\mathrm{ph}, i}=f_{\mathrm{ph},i}\frac{hc}{\lambda_{\mathrm{th},i}},
 	\end{equation}
	where $\lambda_{\mathrm{th},i}$ is the threshold wavelength for the photolysis of species $i$, $h$ is the Planck constant, and $c$ is the speed of light in vacuum. We adopt the photodissociation and photoionization cross sections provided by “PHoto Ionization/Dissociation RATES” (Huebner and Mukherjee, 2015; \url{http://phidrates.space.swri.edu}).
	
	In addition to photolysis reactions, this study considers bimolecular reactions. The bimolecular reaction rate $f_{\rm R}$ can be written as
	\begin{equation}
		f_{\mathrm{R}}=k_{\mathrm{R}}n_{i}n_{j},
 	\end{equation}
	where $n_{i}$ and $n_{j}$ represent the number densities of reactant $i$ and $j$, respectively, and $k_{\rm R}$ is the corresponding reaction rate coefficient. We use the reaction rate coefficients provided by ``The UMIST Database for Astrochemistry 2012" (McElroy et al., 2013; \url{http://udfa.ajmarkwick.net}). We neglect the formation of molecules that have more than one carbon because of low gas densities in the atmospheric regions where the outflow to space accelerates. The energy consumed per unit time by the bimolecular reaction is given by
	\begin{equation}
		q_{\mathrm{R}}=-f_{\mathrm{R}}\Delta E_{\mathrm{R}},
 	\end{equation}
	where $\Delta E_{\mathrm{R}}$ is the heat of reaction, which is positive for endothermic reactions and negative for exothermic reactions. In the evaluation of the heats of reaction, we make use of the enthalpies of formation listed in Le Teuff et al. (2000).  The energy consumption rate $q_{\mathrm{ch}}$ is calculated by summing the consumed amounts of energy by individual chemical reactions including photolysis.

\subsection{Radiative processes}
	In order to calculate the radial profiles of heating rates and photolysis rates in the atmosphere, we model the radiative transfer of parallel stellar photon beams in a spherically symmetric atmosphere by applying the method formulated by Tian et al. (2005a). We consider the XUV absorption by H$_{2}$, H$_{2}$O, H, O, OH, and O$_{2}$. The stellar beam is assumed to be absorbed by the Beer's law. The energy deposition in a given layer along each path is then multiplied by the area of the ring (Fig. 6 in Tian et al., 2005a) to obtain the total energy deposited into the shell segment. The heating rate $q_{\mathrm{abs}}$ is given by the total energy absorbed per unit time by each shell divided by the volume of the shell. We adopt the X-ray and UV spectrum from 0.1 to 280 nm estimated for TRAPPIST-1 by Peacock et al. (2019), and assume that the ratio of the XUV luminosity to the total stellar luminosity is $10^{-3}$ during the pre-main sequence phase.
	
	We consider radiative cooling by thermal line emission of H$_{2}$O, OH, H$_{3}^{+}$, and OH$^{+}$. The radiative cooling rate due to a transition from energy level $i$ to $j$ of radiatively active molecular species $s$ is given by
	\begin{equation}
		q_{\mathrm{rad},s,ij}=n_{s,i}A_{ij}h\nu_{ij}\beta_{ij},
 	\end{equation}
	where $n_{s,i}$ is the population density in level $i$, $A_{ij}$ is the spontaneous transition probability, $h\nu_{ij}$ is the energy difference between level $i$ and level $j$, and $\beta_{ij}$ is the photon escape probability. The total radiative cooling rate by species $s$ is calculated by summing the radiative cooling rates by all the transitions. The population density $n_{s,i}$ is calculated under the assumption of LTE  as
	\begin{equation}
		n_{s,i}=n_{s}\frac{g_{i}}{Z}\mathrm{exp}\left(-\frac{E_{i}}{k_{\mathrm{B}}T}\right)
 	\end{equation}
	where $n_{s}$ and $Z$ are the total number density and partition function of species $s$, and $g_{i}$ and $E_{i}$ are the statistical weight and energy of level $i$. Approximating that half of the photons are emitted outward while the other half is emitted downward and then absorbed by dense, lower atmospheric layers, the bulk escape probability is evaluated by
	\begin{equation}
		\beta(r)=\frac{1}{2}\int_{0}^{\infty}\phi(\nu,r)\,\mathrm{exp}\left[-\int_{r_{\mathrm{top}}}^{r}\alpha_{\rm L}\phi(\nu,r)dr\right] d\nu
	\end{equation}
	where $r_{\mathrm{top}}$ is the radius of the outer boundary (the top of the atmosphere), $\alpha_{\rm L}$ is the integrated absorption coefficient, and $\phi(\nu,r)$ is the line profile, $\alpha_{\rm L}$ is given by
	\begin{equation}
		\alpha_{\rm L}=\frac{c^{2}}{8\pi \nu_{ij}^{2}}n_{s,i}A_{ij}\left(\frac{n_{s,j}g_{i}}{n_{s,i}g_{j}}-1\right).
	\end{equation}
	Assuming the Doppler profile, $\phi(\nu,r)$ is given by
	\begin{equation}
 		\phi(\nu,r)=\frac{1}{\sqrt{\pi}\Delta \nu_{\rm D}}\mathrm{exp}\left[-\frac{1}{(\Delta \nu_{\rm D})^{2}}\left(\nu - \nu_{0}-\frac{u_{s}(r)}{c}\nu_{0}\right)^{2}\right],
	\end{equation}
	where $u_{s}(r)$ is the velocity profile of radiative sources and $\nu_{0}$ is the central frequency of the line profile under no flow; $\Delta \nu_{\rm D}$ is the Doppler width given by
	\begin{equation}
		\Delta \nu_{\rm D}=\frac{\nu_{0}}{c}\sqrt{\frac{3k_{\rm B}T}{m_{s}}},
	\end{equation}
	where $m_{s}$ is the molecular mass of species $s$. The net radiative cooling rate $q_{\mathrm{rad}}$ is obtained by the sum of contributions by H$_{2}$O, OH, H$_{3}^{+}$, and OH$^{+}$. We use the line data provided by HITRAN database (Rothman et al., 2013; \url{http://hitran.org}) and ExoMol databese (Tennyson et al., 2016; \url{http://exomol.com}). We consider 5021 transitions of H$_{2}$O, 383 transitions of OH, 5200 transitions of H$_{3}^{+}$, and 192 transitions of OH$^{+}$ to cover more than 99\% of the total energy emission in the temperature range of 100 - 1000 K.

\subsection{Calculation method}
	The basic equations are solved by numerical time integration until the physical quantities settle into their steady profiles. These equations can be split into advection phases and nonadvection phases. We employ the CIP method to solve the advection phases (Yabe and Aoki, 1991) to keep numerical stability and accuracy for hydrodynamic escape simulations (Kuramoto et al., 2013). After solving the advection phases, we solve the nonadvection phases with a finite-difference approach. The nonadvection phases of the energy equation are solved explicitly. Those of the continuity equations are solved with the semi-implicit method. Those of the momentum equations are solved implicitly using the quantities for the next time step that are obtained by the integration of the energy equation and the continuity equations. 
	
	On radial coordinate, 1000 numerical grids are taken with the grid-to-grid intervals exponentially increasing with $r$. The time step is defined so as to satisfy the CFL condition. For each parameter setup, we continue the time integration until reaching the steady state using the convergence condition described in Tian et al. (2005a).
	
	The upper boundary is set at $r=50\,R_{p}$, where $R_{p}$ is the radius of the planet. The lower boundary is set at $r=R_{p}+1000\,\mathrm{km}$. In each simulation run, the atmospheric density and temperature at the lower boundary are fixed. We assume that only H$_{2}$ and H$_{2}$O exist at the lower boundary. The number density of H$_{2}$ at the lower boundary is set at $1\times 10^{19}\,\mathrm{m^{-3}}$. The number density of H$_{2}$O is given as a parameter in each simulation run. The temperature at the lower boundary is set at 400 K, corresponding to the skin temperature, which is the asymptotic temperature at high altitudes of the upper atmosphere. The other physical quantities at the lower and upper boundaries are estimated by linear extrapolations from the calculated domain. As the initial condition, we use the steady profiles of the pure hydrogen atmosphere and add H$_{2}$O whose number density profiles are given by the hydrostatic structure. The initial velocities of H$_{2}$O is set at $1\times 10^{-5}$ m/s. 
	
\begin{table*}
	\centering
	\caption{Chemical reactions}
	\label{tab:example_table}
	\begin{tabular}{llcll} 
	\hline
	No. & Reaction &  &   & Reaction rate $\mathrm{cm^{3}\,s^{-1}}$\\
	\hline
	R1 & $\mathrm{H}_{2}+h\nu$ & $\to$ & $\mathrm{H}+\mathrm{H}$\\
	R2 &                                        & $\to$ & $\mathrm{H}_{2}^{+}+\mathrm{e}$\\
	R3 &                                        & $\to$ & $\mathrm{H}^{+}+\mathrm{H}+\mathrm{e}$\\
	R4 & $\mathrm{H_{2}O}+h\nu$ & $\to$ & $\mathrm{H}+\mathrm{OH}$\\
	R5 &                                        & $\to$ & $\mathrm{H}_{2}+\mathrm{O(^{1}D)}$\\
	R6 &                                        & $\to$ & $\mathrm{O}+\mathrm{H}+\mathrm{H}$\\		
	R7 &                                        & $\to$ & $\mathrm{OH^{+}}+\mathrm{H}+\mathrm{e}$\\
	R8 &                                        & $\to$ & $\mathrm{O^{+}}+\mathrm{H_{2}}+\mathrm{e}$\\
	R9 &                                        & $\to$ & $\mathrm{H^{+}}+\mathrm{OH}+\mathrm{e}$\\
	R10 &                                        & $\to$ & $\mathrm{H_{2}O^{+}}+\mathrm{e}$\\
	R11 & $\mathrm{H}+h\nu$ & $\to$ & $\mathrm{H^{+}}+\mathrm{e}$\\
	R12 & $\mathrm{O}+h\nu$ & $\to$ & $\mathrm{O^{+}}+\mathrm{e}$\\
	R13 & $\mathrm{OH}+h\nu$ & $\to$ & $\mathrm{O}+\mathrm{H}$\\
	R14 &                                        & $\to$ & $\mathrm{OH^{+}}+\mathrm{e}$\\
	R15 &                                        & $\to$ & $\mathrm{O(^{1}D)}+\mathrm{H}$\\
	R16 & $\mathrm{O_{2}}+h\nu$ & $\to$ & $\mathrm{O}+\mathrm{O}$\\
	R17 &                                        & $\to$ & $\mathrm{O}+\mathrm{O(^{1}D)}$\\
	R18 &                                        & $\to$ & $\mathrm{O^{+}}+\mathrm{O}+\mathrm{e}$\\
	R19 &                                        & $\to$ & $\mathrm{O_{2}^{+}}+\mathrm{e}$\\					
	R20 & $\mathrm{H_{2}}+\mathrm{H_{2}}$ & $\to$ & $\mathrm{H_{2}}+\mathrm{H}+\mathrm{H}$ & $1.0\times 10^{-8}\mathrm{exp}(-84000/T)$\\
	R21 & $\mathrm{H_{2}}+\mathrm{H_{2}O}$ & $\to$ & $\mathrm{OH}+\mathrm{H_{2}}+\mathrm{H}$ & $5.8\times 10^{-9}\mathrm{exp}(-52900/T)$\\
	R22 & $\mathrm{H_{2}}+\mathrm{O_{2}}$ & $\to$ & $\mathrm{O}+\mathrm{O}+\mathrm{H_{2}}$ & $6.0\times 10^{-9}\mathrm{exp}(-52300/T)$\\
	R23 & $\mathrm{H_{2}}+\mathrm{O_{2}}$ & $\to$ & $\mathrm{OH}+\mathrm{OH}$ & $3.16\times 10^{-10}\mathrm{exp}(-21890/T)$\\			
	R24 & $\mathrm{H_{2}}+\mathrm{OH}$ & $\to$ & $\mathrm{O}+\mathrm{H_{2}}+\mathrm{H}$ & $6.0\times 10^{-9}\mathrm{exp}(-50900/T)$\\
	R25 & $\mathrm{H_{2}}+\mathrm{H}$ & $\to$ & $\mathrm{H}+\mathrm{H}+\mathrm{H}$ & $4.67\times 10^{-7}(T/300)^{-1}\mathrm{exp}(-55000/T)$\\
	R26 & $\mathrm{H_{2}}+\mathrm{H_{2}^{+}}$ & $\to$ & $\mathrm{H_{3}^{+}}+\mathrm{H}$ & $2.08\times 10^{-9}$\\
	R27 & $\mathrm{H_{2}}+\mathrm{H_{2}O^{+}}$ & $\to$ & $\mathrm{H_{3}O^{+}}+\mathrm{H}$ & $6.40\times 10^{-10}$\\
	R28 & $\mathrm{H_{2}}+\mathrm{O^{+}}$ & $\to$ & $\mathrm{OH^{+}}+\mathrm{H}$ & $1.70\times 10^{-9}$\\
	R29 & $\mathrm{H_{2}}+\mathrm{OH^{+}}$ & $\to$ & $\mathrm{H_{2}O^{+}}+\mathrm{H}$ & $1.01\times 10^{-9}$\\
	R30 & $\mathrm{H_{2}}+\mathrm{O}$ & $\to$ & $\mathrm{OH}+\mathrm{H}$ & $3.14\times 10^{-13}(T/300)^{2.7}\mathrm{exp}(-3150/T)$\\
	R31 & $\mathrm{H_{2}}+\mathrm{OH}$ & $\to$ & $\mathrm{H_{2}O}+\mathrm{H}$ & $2.05\times 10^{-12}(T/300)^{1.52}\mathrm{exp}(-1736/T)$\\
	R32 & $\mathrm{H_{2}}+\mathrm{O(^{1}D)}$ & $\to$ & $\mathrm{H}+\mathrm{OH}$ & $1.0\times 10^{-9}$\\	
	R33 & $\mathrm{H_{2}O}+\mathrm{H}$ & $\to$ & $\mathrm{OH}+\mathrm{H}+\mathrm{H}$ & $5.80\times 10^{-9}\mathrm{exp}(-52900/T)$\\
	R34 & $\mathrm{H_{2}O}+\mathrm{H^{+}}$ & $\to$ & $\mathrm{H_{2}O^{+}}+\mathrm{H}$ & $6.90\times 10^{-9}(T/300)^{-0.5}$\\
	R35 & $\mathrm{H_{2}O}+\mathrm{H_{2}^{+}}$ & $\to$ & $\mathrm{H_{2}O^{+}}+\mathrm{H_{2}}$ & $3.90\times 10^{-9}(T/300)^{-0.5}$\\
	R36 & $\mathrm{H_{2}O}+\mathrm{O^{+}}$ & $\to$ & $\mathrm{H_{2}O^{+}}+\mathrm{O}$ & $3.20\times 10^{-9}(T/300)^{-0.5}$\\
	R37 & $\mathrm{H_{2}O}+\mathrm{OH^{+}}$ & $\to$ & $\mathrm{H_{2}O^{+}}+\mathrm{OH}$ & $1.59\times 10^{-9}(T/300)^{-0.5}$\\
	R38 & $\mathrm{H_{2}O}+\mathrm{H_{2}^{+}}$ & $\to$ & $\mathrm{H_{3}O^{+}}+\mathrm{H}$ & $3.40\times 10^{-9}(T/300)^{-0.5}$\\
	R39 & $\mathrm{H_{2}O}+\mathrm{H_{2}O^{+}}$ & $\to$ & $\mathrm{H_{3}O^{+}}+\mathrm{OH}$ & $2.10\times 10^{-9}(T/300)^{-0.5}$\\
	R40 & $\mathrm{H_{2}O}+\mathrm{H_{3}^{+}}$ & $\to$ & $\mathrm{H_{3}O^{+}}+\mathrm{H_{2}}$ & $5.90\times 10^{-9}(T/300)^{-0.5}$\\
	R41 & $\mathrm{H_{2}O}+\mathrm{OH^{+}}$ & $\to$ & $\mathrm{H_{3}O^{+}}+\mathrm{O}$ & $1.30\times 10^{-9}(T/300)^{-0.5}$\\
	R42 & $\mathrm{H_{2}O}+\mathrm{H}$ & $\to$ & $\mathrm{OH}+\mathrm{H_{2}}$ & $1.59\times 10^{-11}(T/300)^{1.2}\mathrm{exp}(-9600/T)$\\
	R43 & $\mathrm{H_{2}O}+\mathrm{O}$ & $\to$ & $\mathrm{OH}+\mathrm{OH}$ & $1.85\times 10^{-11}(T/300)^{0.95}\mathrm{exp}(-8571/T)$\\	
	R44 & $\mathrm{H}+\mathrm{OH}$ & $\to$ & $\mathrm{O}+\mathrm{H}+\mathrm{H}$ & $6.00\times 10^{-9}\mathrm{exp}(-50900/T)$\\
	R45 & $\mathrm{H}+\mathrm{H_{2}^{+}}$ & $\to$ & $\mathrm{H_{2}}+\mathrm{H^{+}}$ & $6.40\times 10^{-10}$\\
	R46 & $\mathrm{H}+\mathrm{O^{+}}$ & $\to$ & $\mathrm{O}+\mathrm{H^{+}}$ & $5.66\times 10^{-10}(T/300)^{0.36}\mathrm{exp}(8.6/T)$\\
	R47 & $\mathrm{H}+\mathrm{H^{+}}$ & $\to$ & $\mathrm{H_{2}^{+}}$ & $1.15\times 10^{-18}(T/300)^{1.49}\mathrm{exp}(-228/T)$\\
	R48 & $\mathrm{H}+\mathrm{O_{2}}$ & $\to$ & $\mathrm{O}+\mathrm{O}+\mathrm{H}$ & $6.0\times 10^{-9}\mathrm{exp}(-52300/T)$\\
	R49 & $\mathrm{H}+\mathrm{O_{2}}$ & $\to$ & $\mathrm{OH}+\mathrm{O}$ & $2.61\times 10^{-10}\mathrm{exp}(-8156/T)$\\
	R50 & $\mathrm{H}+\mathrm{O}$ & $\to$ & $\mathrm{OH}$ & $9.90\times 10^{-19}(T/300)^{-0.38}$\\
	\hline
	 & & & & Ref. McElroy et al. (2013)\\
	\end{tabular}
\end{table*}	
	
\begin{table*}
	\centering
	\caption{Chemical reactions}
	\label{tab:example_table}
	\begin{tabular}{llcll} 
	\hline
	No. & Reaction &  &   & Reaction rate $\mathrm{cm^{3}\,s^{-1}}$\\
	\hline
	R51 & $\mathrm{H}+\mathrm{OH}$ & $\to$ & $\mathrm{H_{2}O}$ & $5.26\times 10^{-18}(T/300)^{-5.22}\mathrm{exp}(-90/T)$\\
	R52 & $\mathrm{H}+\mathrm{OH}$ & $\to$ & $\mathrm{O}+\mathrm{H_{2}}$ & $6.99\times 10^{-14}(T/300)^{2.8}\mathrm{exp}(-1950/T)$\\	
	R53 & $\mathrm{O}+\mathrm{H^{+}}$ & $\to$ & $\mathrm{O^{+}}+\mathrm{H}$ & $6.86\times 10^{-10}(T/300)^{0.26}\mathrm{exp}(-224.3/T)$\\
	R54 & $\mathrm{O}+\mathrm{H_{2}^{+}}$ & $\to$ & $\mathrm{OH^{+}}+\mathrm{H}$ & $1.50\times 10^{-9}$\\
	R55 & $\mathrm{O}+\mathrm{H_{3}^{+}}$ & $\to$ & $\mathrm{H_{2}O^{+}}+\mathrm{H}$ & $3.42\times 10^{-10}(T/300)^{-0.16}\mathrm{exp}(-1.4/T)$\\
	R56 & $\mathrm{O}+\mathrm{H_{3}^{+}}$ & $\to$ & $\mathrm{OH^{+}}+\mathrm{H_{2}}$ & $7.98\times 10^{-10}(T/300)^{-0.16}\mathrm{exp}(-1.4/T)$\\
	R57 & $\mathrm{O}+\mathrm{H_{2}O^{+}}$ & $\to$ & $\mathrm{O_{2}^{+}}+\mathrm{H_{2}}$ & $4.0\times 10^{-11}$\\
	R58 & $\mathrm{O}+\mathrm{OH^{+}}$ & $\to$ & $\mathrm{O_{2}^{+}}+\mathrm{H}$ & $7.10\times 10^{-10}$\\	
	R59 & $\mathrm{O}+\mathrm{OH}$ & $\to$ & $\mathrm{O_{2}}+\mathrm{H}$ & $3.69\times 10^{-11}(T/300)^{-0.27}\mathrm{exp}(-12.9/T)$\\
	R60 & $\mathrm{O}+\mathrm{O}$ & $\to$ & $\mathrm{O_{2}}$ & $4.9\times 10^{-20}(T/300)^{1.58}$\\
	R61 & $\mathrm{O(^{1}D)}$ & $\to$ & $\mathrm{O}$ & $1.0\times 10^{-4}$\\
	R62 & $\mathrm{O(^{1}D)}+\mathrm{H_{2}O}$ & $\to$ & $\mathrm{OH}+\mathrm{OH}$ & $2.2\times 10^{-9}$\\
	R63 & $\mathrm{O(^{1}D)}+\mathrm{H_{2}}$ & $\to$ & $\mathrm{H}+\mathrm{OH}$ & $1.0\times 10^{-9}$\\
	R64 & $\mathrm{O(^{1}D)}+\mathrm{O_{2}}$ & $\to$ & $\mathrm{O_{2}}+\mathrm{O}$ & $3.2\times 10^{-11}\mathrm{exp}(70/T)$\\
	R65 & $\mathrm{O(^{1}D)}+\mathrm{O}$ & $\to$ & $\mathrm{O}+\mathrm{O}$ & $8.0\times 10^{-12}$\\		
	R66 & $\mathrm{OH}+\mathrm{H^{+}}$ & $\to$ & $\mathrm{OH^{+}}+\mathrm{H}$ & $2.1\times 10^{-9}(T/300)^{-0.5}$\\
	R67 & $\mathrm{OH}+\mathrm{H_{2}^{+}}$ & $\to$ & $\mathrm{OH^{+}}+\mathrm{H_{2}}$ & $7.6\times 10^{-10}(T/300)^{-0.5}$\\
	R68 & $\mathrm{OH}+\mathrm{O^{+}}$ & $\to$ & $\mathrm{OH^{+}}+\mathrm{O}$ & $3.6\times 10^{-10}(T/300)^{-0.5}$\\
	R69 & $\mathrm{OH}+\mathrm{H_{2}^{+}}$ & $\to$ & $\mathrm{H_{2}O^{+}}+\mathrm{H}$ & $7.6\times 10^{-10}(T/300)^{-0.5}$\\
	R70 & $\mathrm{OH}+\mathrm{H_{3}^{+}}$ & $\to$ & $\mathrm{H_{2}O^{+}}+\mathrm{H_{2}}$ & $1.3\times 10^{-9}(T/300)^{-0.5}$\\
	R71 & $\mathrm{OH}+\mathrm{O^{+}}$ & $\to$ & $\mathrm{O_{2}^{+}}+\mathrm{H}$ & $3.6\times 10^{-10}(T/300)^{-0.5}$\\
	R72 & $\mathrm{OH}+\mathrm{OH^{+}}$ & $\to$ & $\mathrm{H_{2}O^{+}}+\mathrm{O}$ & $7.0\times 10^{-10}(T/300)^{-0.5}$\\
	R73 & $\mathrm{OH}+\mathrm{H_{2}O^{+}}$ & $\to$ & $\mathrm{H_{3}O^{+}}+\mathrm{O}$ & $6.9\times 10^{-10}(T/300)^{-0.5}$\\
	R74 & $\mathrm{OH}+\mathrm{OH}$ & $\to$ & $\mathrm{H_{2}O}+\mathrm{O}$ & $1.65\times 10^{-12}(T/300)^{1.14}\mathrm{exp}(-50/T)$\\	
	R75 & $\mathrm{O_{2}}+\mathrm{H^{+}}$ & $\to$ & $\mathrm{O_{2}^{+}}+\mathrm{H}$ & $2.0\times 10^{-9}$\\
	R76 & $\mathrm{O_{2}}+\mathrm{H_{2}^{+}}$ & $\to$ & $\mathrm{O_{2}^{+}}+\mathrm{H_{2}}$ & $8.0\times 10^{-10}$\\
	R77 & $\mathrm{O_{2}}+\mathrm{H_{2}O^{+}}$ & $\to$ & $\mathrm{O_{2}^{+}}+\mathrm{H_{2}O}$ & $4.6\times 10^{-10}$\\
	R78 & $\mathrm{O_{2}}+\mathrm{O^{+}}$ & $\to$ & $\mathrm{O_{2}^{+}}+\mathrm{O}$ & $1.9\times 10^{-11}$\\
	R79 & $\mathrm{O_{2}}+\mathrm{OH^{+}}$ & $\to$ & $\mathrm{O_{2}^{+}}+\mathrm{OH}$ & $5.9\times 10^{-10}$\\
	R80 & $\mathrm{H^{+}}+\mathrm{e}$ & $\to$ & $\mathrm{H}+\mathrm{PHOTON}$ & $3.5\times 10^{-12}(T/300)^{-0.75}$\\
	R81 & $\mathrm{H_{2}^{+}}+\mathrm{e}$ & $\to$ & $\mathrm{H}+\mathrm{H}$ & $1.6\times 10^{-8}(T/300)^{-0.43}$\\
	R82 & $\mathrm{H_{3}^{+}}+\mathrm{e}$ & $\to$ & $\mathrm{H_{2}}+\mathrm{H}$ & $2.34\times 10^{-8}(T/300)^{-0.52}$\\
	R83 & $\mathrm{H_{3}^{+}}+\mathrm{e}$ & $\to$ & $\mathrm{H}+\mathrm{H}+\mathrm{H}$ & $4.36\times 10^{-8}(T/300)^{-0.52}$\\
	R84 & $\mathrm{O^{+}}+\mathrm{e}$ & $\to$ & $\mathrm{O}+\mathrm{PHOTON}$ & $3.24\times 10^{-12}(T/300)^{-0.66}$\\
	R85 & $\mathrm{O_{2}^{+}}+\mathrm{e}$ & $\to$ & $\mathrm{O}+\mathrm{O}$ & $1.95\times 10^{-7}(T/300)^{-0.7}$\\
	R86 & $\mathrm{OH^{+}}+\mathrm{e}$ & $\to$ & $\mathrm{O}+\mathrm{H}$ & $3.75\times 10^{-8}(T/300)^{-0.5}$\\	
	R87 & $\mathrm{H_{2}O^{+}}+\mathrm{e}$ & $\to$ & $\mathrm{O}+\mathrm{H_{2}}$ & $3.9\times 10^{-8}(T/300)^{-0.5}$\\
	R88 & $\mathrm{H_{2}O^{+}}+\mathrm{e}$ & $\to$ & $\mathrm{O}+\mathrm{H}+\mathrm{H}$ & $3.05\times 10^{-7}(T/300)^{-0.5}$\\
	R89 & $\mathrm{H_{2}O^{+}}+\mathrm{e}$ & $\to$ & $\mathrm{OH}+\mathrm{H}$ & $8.6\times 10^{-8}(T/300)^{-0.5}$\\
	R90 & $\mathrm{H_{3}O^{+}}+\mathrm{e}$ & $\to$ & $\mathrm{H_{2}O}+\mathrm{H}$ & $7.09\times 10^{-8}(T/300)^{-0.5}$\\
	R91 & $\mathrm{H_{3}O^{+}}+\mathrm{e}$ & $\to$ & $\mathrm{O}+\mathrm{H_{2}}+\mathrm{H}$ & $5.60\times 10^{-9}(T/300)^{-0.5}$\\
	R92 & $\mathrm{H_{3}O^{+}}+\mathrm{e}$ & $\to$ & $\mathrm{OH}+\mathrm{H_{2}}$ & $5.37\times 10^{-8}(T/300)^{-0.5}$\\
	R93 & $\mathrm{H_{3}O^{+}}+\mathrm{e}$ & $\to$ & $\mathrm{OH}+\mathrm{H}+\mathrm{H}$ & $3.05\times 10^{-7}(T/300)^{-0.5}$\\	
	\hline
	 & & & & Ref. McElroy et al. (2013)\\
	\end{tabular}
\end{table*}




\end{document}